# Inside the Kin Network: Consanguineous Marriage, Patriarchal Bargaining, and Women's Acceptance of Intimate Partner Violence in Pakistan

**Sana Khalil and Angela Warner**

**Abstract:** While studies on women's acceptance of intimate partner violence (IPV) are extensive, we offer a complementary perspective by examining how marriage and intra-household dynamics influence these attitudes. Using a sample of 15,065 ever-married women from the Pakistan Demographic and Health Survey 2017–18, and applying logistic and ordered logistic regression alongside propensity score and Mahalanobis distance matching, we find a positive association between consanguineous marriage and women's likelihood of accepting IPV. Women's education and empowerment through participation in household decision-making reduce women's acceptance of IPV, while prior experiences of violence are linked to greater acceptance. Moreover, significant regional effects point to the influence of local institutional and normative factors across Pakistan. The association is concentrated in situations that challenge patriarchal control over women's autonomy—going out without permission, refusing sex, or arguing with a husband—rather than domestic tasks such as burning food, and this pattern holds across matching methods. Interpreted through Kandiyoti's patriarchal bargaining framework, these findings identify the marriage system itself as a key institutional factor shaping attitudes toward IPV, and suggest that prevention efforts should extend beyond couples to engage broader kinship networks—in-laws, elders, and extended family—alongside the contextual factors that sustain these norms.

# I. INTRODUCTION

Cross-national evidence consistently identifies South Asia and sub-Saharan Africa as the regions with the highest acceptance of intimate partner violence among women. Tran et al. (2016), analyzing data from 39 low- and middle-income countries, find that South Asia has the highest regional median acceptance rate at 64.9%, with Afghanistan reaching 90.2% — the highest in the study. West and Central Africa follows closely, with a regional median of 57.2% across eight countries including Nigeria, Uganda, Zambia, Ghana, Kenya, Tanzania, Burkina Faso, and Ethiopia, where acceptance rates range from 40% to 80% depending on the scenario and study. La Mattina and Shemyakina (2024), analyzing data from 48 Demographic and Health Surveys across 23 sub-Saharan African countries, find that women who experienced high-intensity armed conflict before age 20 are slightly more likely to accept domestic violence compared to their unexposed peers in the same community. The strongest association occurs with early childhood exposure. The authors suggest this may be partly due to conflict's adverse impact on educational attainment and early age at marriage. Similarly, Sardinha and Nájera Catalán (2018), drawing on Demographic and Health Survey data from 49 low- and middle-income countries, show that regional variation in IPV acceptance maps closely onto structural indicators of gender inequality and human development. Countries with higher Gender Inequality Index scores, lower Human Development Index values, and fewer average years of schooling consistently show higher rates of IPV acceptance among both women and men. South Asia and sub-Saharan Africa — the two regions with the highest IPV acceptance rates globally — also exhibit the highest levels of gender inequality, the lowest educational attainment, and the weakest economic development indicators.

In Pakistan, 42% of women and 40% of men believe wife-beating is justified under at least one circumstance, and these attitudes have shown little improvement over time — women's acceptance has remained stable at 42% across the 2012–13 and 2017–18 PDHS survey cycles, while men's has increased from 34% to 40% (Gallup Pakistan, 2025). A consistent yet underexamined aspect of this literature is the scenario-specific hierarchy of justifications: acceptance peaks in situations threatening autonomy, such as going out without informing the husband or arguing with him, and is lowest in cases like domestic task failures, for example burning food (Waltermaurer, 2012; Sayem et al., 2012; Saud et al., 2021). This pattern appears across various country contexts but has received limited attention. We fill this gap by proposing that consanguineous marriage offers a structurally grounded mechanism: by strengthening kinship networks that enhance surveillance over women's mobility and bodily autonomy, consanguinity specifically increases IPV acceptance in areas where women's behavior is seen as challenging family honor or patriarchal authority, while remaining relatively stable in areas related to routine domestic tasks.

Consanguineous marriage, defined as marriage between close biological relatives, is a widespread kinship practice that affects millions of people across Asia and Africa. It accounts for 20–55 percent of marriages in various countries across North Africa, West Asia, and South Asia (Bittles, 2012) and is found across multiple religious communities, including Muslims, Hindus, Buddhists, Christians, and others (Bittles, 1994). Pakistan has the highest rates of consanguineous marriage globally, with approximately 61 percent of marriages in this category according to recent data. Close-kin marriages have shown remarkable persistence in Pakistan over several decades, declining only modestly from 62 percent in 1990 to 58 percent in 2018. This high prevalence and persistence reflect its deep institutionalization within kinship systems and its enduring significance in shaping everyday social, familial, and economic relations.

Consanguineous marriage operates as a patriarchal institutional structure that consolidates social, economic, and ritual practices within extended families while discouraging marital ties with non-kin through social sanctions such as ostracism, reputational damage, and loss of familial support (Edlund, 2018). By restricting the formation of alliances outside the kin group, consanguinity limits exposure to alternative familial arrangements, gender relations, and norms. Over time, this insularity can generate entrenched belief systems, particularly around gender roles, obedience, and women's responsibilities within marriage. Hence, consanguineous marriage norms not only reproduce kinship ties but also sustain expectations regarding women's behavior, labor, and mobility. In such settings, deviation from prescribed gender roles may be perceived not merely as an individual transgression but as a threat to collective family honor, reinforcing conformity through both explicit and implicit penalties (Khalil, 2024; Khalil and Warner, 2025).

Research on consanguinity highlights a complex set of trade-offs that shape women's attitudes toward patriarchal gender norms. On the one hand, women in consanguineous marriages often experience lower divorce[1] and desertion rates (Lenze and Klasen, 2017; Agha, 2016; Hamamy and Alwan, 2016; Charsley, 2007), reduced dowry obligations (Do et al., 2013; Anukriti and Dasgupta, 2017), and greater familiarity with their spouses and in-laws (Sandridge et al., 2010). Such marriages may also offer greater financial security, as dense kin networks facilitate mutual support during periods of economic hardship (Weinreb, 2008; Shenk et al., 2016). On the other hand, consanguineous marriages are associated with significant disadvantages, including elevated risks of recessive genetic disorders among offspring (Tadmouri et al., 2009; Hamamy, 2012; Kelmemi et al., 2015), a higher incidence of spousal violence (Mobarak et al., 2019), and the reinforcement of patriarchal norms that constrain women's autonomy and decision-making power (Charsley, 2007; Makino, 2019).

Previous literature has consistently linked women's justification of intimate partner violence to their adherence to patriarchal gender norms that prescribe women's submissive social and familial roles (Zaatut and Haj-Yahia, 2016; Tayyab et al., 2017). These attitudes may also reflect indoctrination into rigid gender and spousal roles (for a summary review of the relevant literature, see Haj-Yahia, 1998).

Women in consanguineous marriages may be more likely to rationalize intimate partner violence because consanguinity functions as a patriarchal institution that embeds women more deeply in extended kinship hierarchies, where preserving family honor, cohesion, and reputation is paramount. In such marriages, women are not only wives but also insiders in the husband's kin group, heightening expectations of conformity and loyalty. Reporting or condemning intimate partner violence may be strongly discouraged because it risks exposing private family matters and bringing dishonor to both natal and affinal kin. Norms emphasizing women's responsibility to maintain familial harmony (often framed as moral virtue or duty) can lead women to internalize violence as justified punishment for perceived deviations from gendered expectations. As a result, rationalizing spousal violence[2] becomes a way for women to reconcile personal experiences with dominant norms governing obedience, respectability, and the expectation that women prioritize collective family interests over personal autonomy.

---

[1]Lower observed divorce rates may mask constrained exit options for women, reflecting heightened social and economic barriers that discourage dissolution even in abusive or harmful marital relationships.

[2] In this paper, intimate partner violence (IPV) refers specifically to violence perpetrated by husbands against their wives (spousal violence). Given the Pakistani context, where marriage is the primary socially recognized intimate partnership, the terms IPV and spousal violence are used interchangeably (see Khalil, 2024).

Kandiyoti's (1988) framework of patriarchal bargaining argues that women strategically accommodate patriarchal norms because compliance offers security, social legitimacy, access to family resources, and the prospect of greater authority later in life. In the context of South Asia's "classic patriarchy," where women marry into their husbands' extended families and occupy subordinate positions early in marriage, consanguineous unions may further reinforce these patriarchal bargains by embedding women within dense and overlapping kinship networks. These networks strengthen the monitoring and enforcement of gender norms, particularly regarding women's mobility and bodily autonomy, while reducing opportunities to resist or exit unequal relationships. As a result, deviations from expected gender roles are more likely to be interpreted as violations of familial obligations, increasing the perceived legitimacy of intimate partner violence as a means of enforcing patriarchal norms.

In this vein, Campbell et al. (2024) argue that honor cultures and the propriety norms they impose on women are more likely to emerge in societies where cousin marriage is historically common. Cousin marriage can benefit families through wealth consolidation and tighter kin cooperation. However, it can also create parent–offspring conflict over whom women should marry, with honor codes and punitive sanctions enforcing preferred marriages and deterring "deviations." The authors use Demographic and Health Survey data from Jordan, Egypt, Pakistan, and Turkey, estimating logistic regression models to examine regional variation. They find that regions with higher rates of first-cousin marriage have greater odds of justifying wife beating for going out without telling the husband, and lower odds of women's employment outside the home (two proxies for honor-related control).

Elsewhere, a major strand of the literature has used large cross-national surveys to show that IPV acceptance is widespread across low- and middle-income countries, including Asia (Waltermaurer, 2012). Tran et al. (2016), analyzing 39 low- and middle-income countries, found substantial acceptance of IPV among both women and men and showed that such attitudes are closely tied to structural gender inequality rather than to individual characteristics alone. They found that women's acceptance of IPV varies enormously between countries, from around 2% of women in Argentina to over 90% in Afghanistan; among men, acceptance ranged from about 5% in Belarus to nearly 75% in the Central African Republic. Acceptance is generally higher in African and South Asian countries and lower in Central/Eastern Europe, Latin America, and the Caribbean.[3] Across different regional settings, acceptance of IPV is more prevalent among younger respondents, those with less education, rural residents, and households in the lowest wealth quintile. This reinforces the idea that acceptance of IPV is not country-specific but is closely tied to structural factors. Similarly, Sardinha and Nájera Catalán (2018), using data from 49 low- and middle-income countries, found that on average, women are more likely than men to justify IPV, but this pattern reverses in some countries, e.g., Ukraine and Armenia, where men show higher acceptance. Their pooled global estimate found women are 1.39 times more likely than men to justify intimate partner violence (95% CI 1.37–1.41). Their analysis also showed that higher societal gender inequality, discriminatory social institutions, and weaker governance are associated with greater acceptance of IPV. Contextual factors such as rural residence, economic development, and exposure to conflict also relate to attitudes.

[3] Tran et al.'s (2016) analysis of pooled data from demographic and health surveys (DHS) across 39 low- and middle-income countries showed that men are more likely than women to justify IPV in Central and Eastern Europe. Conversely, in most countries in Asia and Africa, *women* are more likely than men to endorse justifications for IPV.

A number of country-level studies consistently identify women's education, household wealth, and economic independence as strong protective factors against women's acceptance of IPV (Tran et al., 2016; Biswas et al., 2017; Lassi et al., 2021; Shreemoyee et al., 2025). For example, Biswas et al. (2017) used nationally representative data from the Bangladesh Demographic and Health Survey 2007 for ever-married women aged 15–49 to examine the correlates of women's acceptance of IPV. They found that women in the wealthiest households were about 40% less likely to justify violence compared to the poorest, while highly educated women were approximately 50% less likely than women with no education. Husband's education had a similar but somewhat weaker effect. Women who married before age 18 were around 20% more likely to accept IPV, and women whose household decisions were made by others, rather than jointly or independently, were about 14–40% more likely to justify violence depending on the survey year. Similar results were found in a small-scale study in Dhaka, Bangladesh, in which Sayem et al. (2012) surveyed 331 Bangladeshi women in five disadvantaged areas of Dhaka city, using a shortened version of the Inventory of Beliefs about Wife Beating (IBWB) scale to measure attitudes. They found that seven factors—respondent's education, husband's education, media exposure, current age, age at marriage, recent physical violence, and micro-credit receipt—together accounted for just under 43% of the variation in women's attitudes toward justifying IPV. Greater justification correlated with older age and later marriage age, whereas lower justification was linked to higher respondent and husband's education, more media exposure, recent physical violence experience, and receiving micro-credit.

Other researchers have also reported conflicting findings on how education and poverty shape women's acceptance of IPV. For example, Jesmin (2017) examined community-level effects on married women's views of intimate partner violence in Bangladesh, using multilevel regression on data from 16,480 women across 600 communities in the 2011 Bangladesh Demographic and Health Survey. While some of Jesmin's findings align with the general literature—such as greater education, household wealth, decision-making power, and media exposure reducing women's likelihood of justifying IPV—others contradict this pattern. Specifically, after accounting for individual factors, Jesmin found that women in poorer communities with lower literacy rates were *less* likely to justify IPV. This challenges the social disorganization theory, which posits that concentrated disadvantage increases IPV tolerance. She proposes that severe material hardship may weaken patriarchal social structures, reduce men's authority, and give women more freedom to reject abuse. Importantly, Jesmin identified community norms about IPV against wives as the strongest predictor of an individual's acceptance of IPV. This aligns with Sardinha and Nájera Catalán's (2018) finding that discriminatory social institutions and societal gender inequality influence acceptance more than individual characteristics. This effect is likely to be especially pronounced in countries like Pakistan, where communal norms are strong. For example, Nadeem and Malik (2021), using data from the Punjab province in Pakistan, show that women residing in communities where wife-beating is socially accepted are more than three times as likely to justify it across all five standard dimensions.[4]

These regional patterns are especially well-documented in Pakistan. Nasrullah et al. (2017) used data from the Pakistan Demographic and Health Survey (PDHS) 2012–13 to show that child marriage is associated with greater justification of IPV among young married women. Similarly,

---

[4] The five standard dimensions, drawn from the Demographic and Health Survey module on attitudes toward intimate partner violence, ask whether a husband is justified in hitting or beating his wife if she: goes out without informing him, neglects the children, argues with him, refuses to have sexual intercourse with him, or burns the food.

Saud, Ashfaq, and Mas'udah (2021) also used data from the PDHS 2012–13 to show that women's acceptance of IPV varies significantly across sociodemographic characteristics. Higher education and greater household wealth are the strongest and most consistent negative predictors of acceptance, with more educated and wealthier women less likely to justify wife beating. Older women are also less likely to justify it. Regional and urban–rural differences are significant, reflecting the heterogeneity in gender norms across Pakistan's administrative regions. In terms of scenario-specific patterns, arguing with the husband is the most widely accepted justification for wife beating, while burning food is the least accepted. However, the drivers of these scenario-specific patterns remain unexplored, a gap this study attempts to address.

Other studies from Pakistan emphasize the roles of familial and community-level norms, as well as women's socioeconomic empowerment, in shaping women's attitudes towards IPV (Nadeem and Malik, 2021; Lassi et al., 2021). Raza and Pals (2025) analyzed data from two waves of the (PDHS)—2012–13 and 2017–18—using weighted binary logistic regression models that included interaction effects. About 47% of women in the combined sample justified intimate partner violence (IPV) for at least one reason. Consistent with most South Asian studies, women showed more acceptance of IPV than men. Importantly, the authors highlight the impact of "patriarchal surveillance"—the social pressure that compels women to adhere to expected norms—on women's reported acceptance of IPV. For interviews, the authors categorized women into two groups: (1) those for whom no one was present or, if present, was not listening during the interview; and (2) those who were overheard, with another family member (husband, child, or female relative) listening in. Their results show that in the full, undivided sample, women's overall tolerance appears higher than men's and has increased over time. But once the authors separate women who were overheard by a family member during the interview from those who were not, the picture changes substantially. Women who were not overheard showed stable attitudes across the five-year period and levels similar to those of men, while women who were overheard displayed sharply elevated tolerance (92% higher odds) and increasing tolerance over time.

Overall, the cross-country empirical literature highlights three key insights about women's acceptance of IPV. First, this acceptance is widespread enough to be considered a central aspect of gender inequality, rather than a marginal belief specific to individuals, countries, or cultures. This topic warrants immediate policy-related research to inform context-specific solutions for healthcare and community practitioners. Women's attitudes toward IPV are also directly linked to the United Nations' SDG 5 (Sustainable Development Goal aimed at achieving gender equality and empowering all women and girls by 2030), which includes eliminating all forms of violence against women and girls and eradicating harmful practices. Second, factors such as women's education and economic independence consistently help reduce women's acceptance of IPV. Third, women's attitudes toward IPV are deeply rooted in structural factors: they are influenced more by cultural norms, poverty, community standards, and patriarchal institutions than by individual opinions alone.

The analysis of the relevant literature shows that while marriage dynamics, often captured in regression models by women's age at marriage, are frequently examined as predictors of attitudes toward IPV, their role in mediating broader familial and kinship norms remains underexplored. These norms impact women's agency and beliefs, often through communal rewards and punishments. A substantial number of studies on women's acceptance of IPV have predominantly focused on identifying socio-economic and demographic correlates rather than the role of the marriage system and kinship structures. Specifically, very few have explored how consanguineous marriage dynamics can influence women's attitudes toward IPV. Moreover, a significant majority of studies examining women's acceptance of IPV have relied on logistic regression, which restricts causal interpretation because it does not account for non-random selection into key factors like education, timing of marriage, and employment.

Propensity score matching (PSM) provides a valuable alternative by enhancing comparability between treatment and control groups based on observable traits (Rosenbaum and Rubin, 1983). Through matching individuals with similar socioeconomic and demographic profiles, PSM minimizes selection bias and offers more reliable estimates of how specific factors—such as education—influence attitudes toward IPV. In this setting, PSM approximates a counterfactual, allowing researchers to evaluate how women's acceptance of IPV could vary if similarly situated individuals experienced different levels of exposure. Nonetheless, although PSM improves causal inference over traditional regression, it is limited to observed variables and cannot address unmeasured factors like deep-seated social norms or intra-household dynamics (Stuart, 2010).

Our study offers three main contributions to the existing research. First, it adds to the literature by examining how kinship dynamics (as captured through consanguinity) shape women's attitudes toward IPV. While most previous research treats households as uniform units, our study emphasizes that marriage within kin networks can heighten social oversight or patriarchal surveillance, reinforce patriarchal norms, and distinctly shape women's views on violence. From this perspective, our study provides a deeper understanding of how patriarchal control is exercised through family structures and marriage dynamics, beyond individual or community factors. Second, our research investigates the reasons behind scenario-specific patterns observed in women's acceptance of IPV. The literature consistently documents that scenario-specific patterns exist — burning food always attracts the lowest acceptance, going out without permission, refusing sex, and arguing always attract the highest. However, to our knowledge, no study has explored the specific question: why is acceptance of IPV concentrated in autonomy-threatening scenarios (going out, refusing sex) rather than perceived domestic task failures, such as burning food? Our study addresses that gap by proposing and testing a kinship-based mechanism: consanguineous marriage selectively reinforces acceptance of IPV in domains directly tied to patriarchal control over women's mobility and bodily autonomy.

Finally, our study advances methodology by employing propensity score matching (PSM) to address the selection bias in studying how consanguineous marriage can influence women's attitudes towards IPV. Since marriage type is not assigned randomly and is influenced by socio-economic factors, region, and cultural norms, PSM offers more reliable comparisons between women in consanguineous and non-consanguineous marriages. Logistic regression conditions on observed covariates within the outcome model but does not account for the non-random selection into consanguineous marriage itself. Women who marry within their kin group differ systematically from those who do not in ways that also shape attitudes toward IPV — such as education, wealth, and regional background — and simply including these variables as controls does not address the underlying selection process. Propensity score matching explicitly models the probability of being in a consanguineous marriage given observed characteristics, constructing a comparable control group before outcomes are compared, and thereby offering more credible estimates of the independent association between marriage type and IPV acceptance. Thus, propensity score matching restricts outcome comparisons (whether they accept IPV or not) to women with similar observable characteristics, ensuring that the treatment group (women in consanguineous marriages) and the control group (women not in consanguineous marriages) are comparable based on their socio-demographic characteristics.

## II. Country Context

Understanding the socioeconomic landscape of Pakistani women is essential for contextualizing their attitudes toward intimate partner violence. The same structural conditions that limit women's economic independence and autonomy are precisely those that the literature identifies as the strongest predictors of higher IPV acceptance.

Pakistan ranks last among 148 countries in the Global Gender Gap Report (World Economic Forum, 2025), reflecting a deeply patriarchal society where gender inequality is systematically reinforced in economic, cultural, and institutional spheres (Sarwar and Farid, 2024). The traditional division of roles assigns women to domestic and caregiving responsibilities while positioning men as breadwinners, a structure that deeply shapes women's participation in economic and public life in Pakistan (Zulfiqar et al., 2026). This structure is deeply ingrained, influencing women’s access to work and education, their decision-making power within families, and their acceptance of intimate partner violence (IPV) (Nadeem and Malik, 2021; Malik et al., 2024; Saud et al., 2021).

Women's role in Pakistan's economy both mirrors and sustains their wider social subordination. Female labor force participation stands at just 24%—well below both the male rate of 80% and the global female average of 49%— ranking among the world's lowest (World Bank, 2025). Regionally, South Asia's female labor force participation averages 33% (World Bank, 2025). Of employed women, 66% work in agriculture (World Bank, 2023), and about 78% are in the informal sector (UN Women, 2016). Over half (54%) of female employment is unpaid or informal family labor, limiting women's financial independence and increasing their risk of exploitation (Khalil and Warner, 2025). Pakistani women experience a 34% gender wage gap, substantially above the global average of 20%, and make up nearly 90% of the country's bottom 1% of earners (International Labour Organization, 2019).

These patterns also reflect cultural norms and social institutions that systemically restrict women's participation in economic life. Women's economic independence and employment are widely stigmatized in Pakistan, with negative attitudes operating at both the household and community levels (Hussain and Jullandhry, 2020) as well as in the labor market, where employers perceive women as inferior to men in terms of productivity, punctuality, and professionalism (Khalil, 2025). Moreover, women's decisions regarding education and employment are frequently determined by male relatives rather than reflecting their own autonomous preferences, and gender norms form documented barriers to women's participation in paid work (Sarwar and Imran, 2019).

The link between women’s education and empowerment may not be straightforward in the Pakistani context. For example, Zulfiqar et al. (2026), drawing on qualitative interviews with highly educated women in Pakistan, portray empowerment as a negotiated, context-dependent process rather than a straightforward outcome of education. Women often exercise “reformist agency,” deliberately conforming to societal norms—such as adhering to dress codes and fulfilling family expectations—to access education, employment, and mobility. This balancing act enables them to steadily expand opportunities for themselves and other women. Empowerment and agency cannot be understood in isolation from social, cultural, and familial contexts. This findings suggest that policies focused solely on gender parity in education are insufficient; meaningful empowerment requires addressing the broader structural and normative constraints that shape women's lives.

Violence against women is a significant public health and human rights issue in Pakistan, affecting women across all socioeconomic backgrounds and representing one of the most direct expressions of patriarchal control (Amir-ud-Din et al., 2021; Nadeem and Malik, 2021). Using data from the PDHS 2012–13, LaBore et al. (2021) found that approximately 20% of ever-married women reported experiencing physical violence from their husbands in the past year. Emotional violence was reported at even higher rates, with 28% experiencing some form in the past year. Overall, 32.6% of women in the same sample experienced at least one form of physical or emotional intimate partner violence in the past year. A separate study using the same data estimated the overall prevalence of emotional and physical violence at 36.4% and 18.4%, respectively (Iqbal and Fatmi, 2021).

Rates of intimate partner violence are shaped by socioeconomic and demographic factors. Most studies find that physical or emotional violence is more prevalent in rural areas (37.4%) than in urban areas (27.0%), and is highest among women with no formal education and among those aged 30–39 (LaBore et al., 2021). Higher rates of violence are also associated with husbands who are uneducated, unemployed, drink alcohol, or are over 45 years old, as well as with households experiencing greater poverty (Iqbal and Fatmi, 2021). Among women who attained only a primary-level education, 82.7% reported spousal violence, compared to just 4.1% of women with higher education (Amir-ud-Din et al., 2021). Intergenerational exposure also plays an important role: women whose fathers hit their mothers were substantially more likely to experience spousal violence themselves, reflecting the transmission of both behavioral norms and tolerance for violence across generations.

Despite these high prevalence rates, studies from Pakistan show that domestic violence is widely normalized, and women commonly justify intimate partner violence at both personal and societal levels. For example, Amir-ud-Din et al. (2021) use data from the PDHS 2012–13 and find that while 55% of women surveyed did not condone violence against women under any circumstances, 15% believed violence was justified in all five standard DHS situations. Additionally, the socioeconomic factors influencing acceptance are similar to those associated with experiencing violence. Women with secondary or higher education have significantly lower odds of accepting domestic violence than those with no education or only primary schooling (Akram, 2021). The husband's education also has a significant effect, though comparatively smaller than the effect of the woman's own education (Akram, 2021; Amir-ud-Din et al., 2021). As household wealth increases, women's odds of accepting domestic violence decrease consistently, with the lowest odds of acceptance found among women in the wealthiest households (Akram, 2021; Nadeem and Malik, 2021). Other factors that decrease women's acceptance of intimate partner violence include marrying after age 18, greater empowerment and decision-making authority, and exposure to print or digital media (Akram, 2021). On the other hand, Akram (2021) uses data from the 2017–18 PDHS and finds significantly higher odds of accepting domestic violence among women in consanguineous marriages.

These patterns of IPV prevalence and acceptance in Pakistan are not unique to the country but reflect broader regional trends. Research using Multiple Indicator Cluster Surveys (MICS) data from Afghanistan, Bhutan, Nepal, and Pakistan (Punjab and Sindh provinces) found that, overall, 58.8% of women across the four countries agreed that IPV is justified for at least one reason, though prevalence varied widely from 44.2% in Punjab, Pakistan, to 94% in Afghanistan (Shrestha et al., 2022). A study by Pettitt et al. (2024) using MICS data from Bangladesh, Laos, and Nepal found that approximately 27.8% of ever-married women across the three countries accepted IPV, with no statistically significant differences in IPV acceptance. On this basis, the authors suggest that predictors of IPV acceptance may broadly apply across the South Asian region. Education was the most influential factor at both national and regional levels, with higher educational attainment reducing IPV acceptance. Increased wealth was also associated with lower IPV acceptance. The study found no significant association between rural or urban residence and women's IPV acceptance, which diverges from most country-level research in Pakistan.

Overall, this section situates women's acceptance of IPV in Pakistan within a broader cultural, regional, and global context Across these settings, IPV prevalence and acceptance are shaped by structural and institutional factors—including patriarchal norms, constrained access to education and economic resources, household poverty,

## III. Selected literature and Theoretical framework

Women's attitudes toward intimate partner violence (IPV) can be conceptualized as the outcome of interacting institutional, economic, and cultural forces embedded within patriarchal social structures. Rather than viewing attitudes toward IPV as purely individual preferences, this framework treats them as socially produced and reinforced through kinship systems, household power relations, economic dependence, and normative environments. In this regard, existing studies show that acceptance of IPV is rooted in patriarchal systems that legitimize male authority and women's subordination (Madhani et al., 2017; Mondal and Paul, 2021). Within these systems, violence functions as a mechanism of coercive control, enforcing compliance with rigid gender roles related to obedience and sexuality (Campbell et al., 2024), as well as mobility and domestic labor (Khalil, 2024). Women's justification of IPV—particularly in response to perceived violations of domestic gender norms (e.g., arguing with the husband, refusing intercourse, neglecting conventional household duties)—reflects the internalization of these patriarchal expectations rather than approval of violence per se (Haj-Yahia, 1998; Tayyab et al., 2017).

Waltermaurer's (2012) cross-national review of attitudes toward IPV across 67 countries found that some form of justification is universal, though rates vary widely. The review also showed that women justify IPV at higher rates than men in about two-thirds of the studies, with some scenarios showing gender gaps as high as 35 percentage points. Multivariate analyses across nine studies included in the review identified predictors of greater IPV justification, including younger age, lower education, poverty, unemployment, rural residence, and patriarchal household structures. In cross-country comparisons, less than 1% of respondents in New Zealand and Cyprus justified IPV in any scenario. Conversely, among 67 countries mostly in sub-Saharan Africa and the Middle East—including Nigeria, Uganda, Zimbabwe, Kenya, and Jordan—up to half or more justified IPV. The most common reasons cited were a woman neglecting her children or leaving without her husband's consent. Although poor home care was the least frequent justification overall, it was the most common factor in Jordan (Waltermaurer, 2012).

Household power relations influence how norms translate into attitudes and experiences of violence. Evidence from India and Pakistan consistently shows that women's decision-making power, control over resources, education, and wealth reduce both the likelihood of experiencing IPV and the acceptance of violence (Mondal and Paul, 2021; Malik et al., 2024; Tayyab et al., 2017). Conversely, husbands' controlling behaviors and women's acceptance of spousal violence increase women's risk of IPV, highlighting the interaction between material dependence and ideological subordination. Access to outside financial resources can help shift these patriarchal attitudes. For example, Mitra et al. (2021) explore whether access to remittance income influences women's acceptance of domestic violence in Pakistan, focusing on Punjab province. Their findings show that women in remittance-receiving households are about 7 percentage points less likely to justify domestic violence than women in similar non-recipient households. These results suggest that income gains and increased household resources can weaken some patriarchal attitudes. However, the effects vary across different contexts. When broken down by scenario, the decline in acceptance of violence mainly appears in cases involving refusal of intercourse and burning food. In contrast, attitudes toward violence in other situations (such as women's mobility without permission from their husbands or perceived neglect in child-rearing) tend to be more resistant to change. This variation shows that while access to remittances and financial independence may improve women's bodily autonomy in specific areas, deeply rooted gender norms governing other aspects of marriage may persist. Cross-national and country-specific evidence further indicates that attitudes toward intimate partner violence are heavily influenced by sociocultural and socioeconomic factors. Using nationally representative data from Afghanistan, Akbary et al. (2022) find that higher education levels and greater household wealth are consistently linked to lower acceptance of wife beating among both men and women, while ethnicity and place of residence create significant differences in attitudes. Their findings also reveal gendered pathways: older women and those living in urban areas are less likely to justify domestic violence, whereas frequent media exposure increases acceptance among men but not women.

Cultural norms, stigma, and lack of institutional support can also shape women's acceptance of and responses to intimate partner violence. Madhani et al. (2017), for example, provide qualitative and quantitative evidence on women's perceptions and responses to domestic violence in Pakistan using data from a community-based observational study of 1,325 women receiving antenatal care in Hyderabad. Their findings revealed that while nearly half of respondents identified physical violence as a form of violence against women, far fewer extended that recognition to non-physical forms (such as verbal abuse, controlling behavior, conflict with in-laws, or excessive domestic burdens), and only a negligible share identified sexual abuse as such. These patterns suggest that patriarchal norms and women's subordinate social position normalize domestic abuse and constrain women's responses, even when violence is recognized as harmful. At the same time, women's recognition of physical abuse as violence suggests that tolerance does not necessarily reflect endorsement, but rather constrained agency within rigid social and familial structures.

A qualitative study by Shams et al. (2020) offers insight into married women's views on domestic violence in Ahvaz, Iran. This study involved focus group discussions with 30 participants recruited from healthcare facilities and identified five key findings. First, most participants recognized that domestic violence is widespread in Ahvaz, linking it to cultural traits, poverty, ethnicity, and climate stress. Second, they viewed violence as encompassing more than physical abuse, including restrictions on education, employment, and expression. Third, contributing factors were categorized as behavioral—such as women's limited knowledge of legal rights and what constitutes violence—and nonbehavioral, mainly economic dependence on husbands, which was seen as the main reason women tolerate and stay in violent situations. Fourth, participants suggested raising awareness through mass media, training, and community education. Fifth, they emphasized the need for stronger law enforcement, better legal protections for abused women, and access to employment and income opportunities to reduce women's vulnerability.

Ahmad et al. (2021) frame spousal violence as a product of interacting factors operating at multiple levels of the social environment, rather than solely within individual households. Using

National Family Health Survey (2015–2016) data for 66,013 ever-married women in India, they estimate multilevel logistic regression models and show that IPV risk is shaped not only by individual and household characteristics (e.g., age, education, parity, wealth, social group) but also by contextual conditions at the community, district, and state levels. In particular, community-level norms (measured by the share of men and women who justify wife beating) are strongly associated with higher odds of women's reported spousal violence, alongside broader macro-context proxies such as district crime rates and state prevalence of underage marriage. This social–ecological evidence supports theorizing IPV attitudes and experiences as outcomes of nested gender regimes, in which household dynamics are conditioned by community norms and wider institutional environments. In this vein, Sardinha and Nájera Catalán (2018) provide cross-national evidence, using Demographic and Health Survey data from 49 low- and middle-income countries, that attitudes toward domestic violence are deeply embedded in broader social, economic, and political structures, rather than reflecting individual beliefs alone. Drawing on data from 49 countries over 2005 and 2017, the authors show that 36% of the respondents across these countries justified intimate partner violence against women in at least one circumstance, with prevalence ranging dramatically from 3% in the Dominican Republic to 83% in Timor-Leste. Women were more likely than men to justify IPV in 36 of 49 countries, particularly in Sub-Saharan Africa and South/Southeast Asia. Men were more likely than women to justify IPV in Latin America, the Caribbean, and Central/West Asia. Using stepwise multiple linear regression, the authors identified distinct gendered patterns in country-level predictors. Higher female literacy rates were the strongest mitigating factor against women's acceptance of IPV, while men's acceptance was more influenced by political structures, with lower justification found in more democratic nations. Both women and men showed higher acceptance of IPV in countries experiencing political conflict and those with limited economic rights for women. Paradoxically, countries with greater representation of women in national parliaments showed higher levels of acceptance of IPV among women and wider gender gaps. The authors suggest that where women's parliamentary representation is driven by gender quotas, it may be endogenous to high-inequality contexts: quotas are often adopted precisely where patriarchal norms remain entrenched**,** and women's representation may therefore be more symbolic than substantive, reflecting limited institutional power rather than broad-based empowerment.

Women's acceptance of IPV may reflect a strategic adaptation to patriarchal constraints, rather than an unconditional endorsement of male authority or violence. In this regard, some researchers have shown that these attitudes vary across domains and may respond to different mechanisms. For example, Mitra et al.'s (2021) evidence on remittances is particularly valuable in showing that empowerment interventions can shift attitudes in some domains (such as bodily autonomy around intercourse and food) but not others (freedom of mobility and childcare practices). This suggests that the scenario-specific structure of IPV acceptance reflects distinct underlying mechanisms.

Yount et al. (2013) conducted a survey experiment in rural Bangladesh, demonstrating that women's reported acceptance of intimate partner violence varies significantly based on the context of a wife's behavior. When the wife's actions were described as unintentional, justification for wife beating was very rare, with only 1–8% of women endorsing violence and just 9% justifying it under any circumstances. Conversely, when the same behaviors were framed as deliberate violations of gender norms, acceptance rose sharply—ranging from 40% to 70%, depending on the scenario, and 80% of respondents justified violence in at least one case. Cognitive interviews showed that some women personally opposed wife beating but still answered that it was 'justified,' believing this was socially expected or feeling limited by patriarchal institutions. Therefore, reported acceptance does not necessarily reflect personal approval of violence. The authors also argue that not all norm violations carry the same social meaning. Public acts that threaten family authority or honor (e.g., disobeying elders, arguing with husbands) attract stronger justification than more private disputes.

These findings raise an important question: why do some women continue to endorse or express support for violence even when they may privately reject it? Drawing on Kandiyoti's (1988) framework of patriarchal bargaining, Sardinha and Nájera Catalán (2018) interpret women's higher acceptance of intimate partner violence in many low- and middle-income countries as a strategic response to structural constraints. In highly patriarchal contexts, women may both be pressured and strategically motivated to conform to gendered norms that legitimize violence against women. Such internalization does not necessarily reflect genuine approval of IPV, but rather may function as a coping or survival strategy. By justifying IPV, women may reduce personal blame, preserve moral or social legitimacy, signal conformity to expectations of being a “good wife,” or mitigate the risk of further violence or social sanction.

This conceptual framework, as outlined by the literature above, rejects methodological individualism and positions attitudes as outcomes of nested gender regimes, operating across household, community, district, state, and national levels. Kandiyoti’s (1988) framework can also be invoked to understand why women may justify IPV.

Kandiyoti argues that women are not passive victims of patriarchy but active — if structurally constrained — agents who develop strategies to maximize their security and minimize their disadvantages within whatever patriarchal system governs their lives. She calls these strategies "patriarchal bargains." Women navigate the rules of the patriarchal game rather than simply submitting to or resisting it, and in doing so they often appear to endorse or reproduce the very norms that subordinate them.

In classic patriarchy, prevalent in sub-Saharan Africa, the Middle East, and parts of South and East Asia, societies are marked by patrilineal kinship, patrilocal residence, and collective male control over women, often exercised by extended families led by men. Women typically follow a defined life course: marrying young with low status, subordinate to mothers-in-law and elder women, but gradually gaining influence and stability as they have sons and eventually become mothers-in-law themselves. Women’s acceptance of these norms—marked by deference, subordination, and limited mobility—serves as a trade-off for long-term security within the family.

Kandiyoti's (1988) framework of patriarchal bargaining informs this paper in two interrelated ways, both of which are operationalized through the lens of consanguinity. First, the framework helps explain why women embedded within overlapping natal and marital kin networks — as is characteristic of consanguineous marriages — may strategically justify IPV. Through kinship consolidation, consanguineous marriages reinforce patrilineal authority and collective control over women's productive and reproductive capacities. When women perceive deviation from patriarchal gender norms as a threat to the kinship-based security they depend upon, violence comes to be rationalized as legitimate discipline rather than recognized as abuse (Akbary et al., 2022; Khalil, 2024). Accepting IPV in this context functions as a survival strategy: it reduces personal blame, signals conformity to family expectations, preserves access to lineage-based resources, and protects women's position within the kinship hierarchy. Second, the framework clarifies the scenario-specific pattern in women's acceptance of IPV that this paper documents empirically. When women assert spatial mobility or refuse sexual intercourse, they challenge the terms of the patriarchal bargain most directly — precisely the domains where overlapping kin networks have the greatest stake in enforcing compliance. Violence in these contexts is rationalized as enforcement of gendered expectations. By contrast, perceived domestic failures such as burning food do not threaten the core of the bargain in the same way, consistent with the substantially lower acceptance rates observed in that scenario. This ambiguity is compounded by the dual nature of consanguinity's effect on women's fallback positions: proximity to natal kin may in some settings strengthen women's outside options, while in others — particularly when combined with extended family residence — it may instead constrain exit options and intensify conformity pressures. Women with weaker fallback positions are thus more prone to endorse norms that justify IPV. The extent of kinship consolidation—measured in this paper by household size and women's participation in selecting their spouses—influences the strength of this mechanism.

Based on this conceptual framework, the next section empirically examines whether and how consanguineous marriage is associated with women's acceptance of intimate partner violence (IPV) in Pakistan. Specifically, this study tests the following hypotheses:

**Hypothesis 1 (IPV incidence):**
Compared to women in non-kin marriages, women in consanguineous marriages are more likely to accept intimate partner violence.

**Hypothesis 2 (scenario-specific attitudes in acceptance of IPV):**
Compared to women in non-kin marriages, women in consanguineous marriages are more likely to justify intimate partner violence in domains that directly challenge patriarchal control over women's autonomy — specifically, their freedom of movement and bodily autonomy — than in domains linked to routine domestic performance.

## IV. DATA AND METHODS

To test the link between consanguineous marriage and women's acceptance of IPV, we use nationally representative survey data from the Pakistan Demographic and Health Survey (PDHS) 2017–18, conducted by the Pakistan Bureau of Statistics, the National Institute of Population Studies, and ICF.[5] The sample consists of 15,065 ever-married women aged 15–49. The survey provides detailed information on marriage type, employment, household decision-making, and attitudes toward spousal violence, along with detailed demographic and socioeconomic controls.

Table 1 provides descriptive statistics for the variables of interest. The primary explanatory variable is consanguineous marriage, which equals 1 if the respondent reports any blood relationship with her husband, including first cousins, second cousins, or other relatives. The outcome of interest is women's acceptance of intimate partner violence. Control variables include age, education, husband's education, household wealth, urban/rural residence, household size (number of household members), wife's say in choosing husband, past violence experience, and region fixed effects.

In our analysis of the PDHS sample, we observe that consanguineous marriages account for 9,381 of 15,065 observations (62%), highlighting the prevalence of close-kin marriages in Pakistan. Women in these marriages consistently show higher IPV acceptance rates. The largest differences are seen in going out without permission (32.3% vs. 39.9%) and refusing sexual intercourse (27.9% vs. 33.8%), while the smallest difference occurs with burning food (17.1% vs. 20.9%), which also has the lowest overall acceptance—reflecting its status as the least justifiable reason for violence.

---

[5] DHS 2017–18 is the fourth survey that was conducted in Pakistan. The first demographic health survey was conducted in 1990-91. The second DHS was published in 2006–07, while the third one was conducted in 2012–13. While the 2017–18 PDHS is the most recently available wave at the time of writing, a fifth PDHS survey is anticipated, and future research can examine whether the patterns documented here persist or have shifted over time.

The educational gap between the two groups of women is substantial. Women in consanguineous marriages have considerably less education, averaging 4 years of schooling, compared to 5.4 years for women in non-kin marriages. A similar, though smaller, difference exists for husbands—averaging 6.8 years in consanguineous marriages versus 7.5 years in non-consanguineous marriages. Education is a key protective factor, as shown in previous research (see Waltermaurer, 2012, for a review of this literature). Wealth distribution follows a similar pattern: women in consanguineous marriages are disproportionately represented in the poorest wealth quintiles, with 44.3% in the bottom two quintiles, compared to 34.7% of women in non-kin marriages. Consanguineous marriages are also more prevalent in rural areas. In the PDHS sample, 55% of women in consanguineous marriages reside in rural areas, compared with 46.7% of women in non-kin marriages.

An important insight from the descriptive analysis concerns women's reported experiences of violence. The three indicators of past violence—experience of less severe physical violence, severe physical violence, and sexual violence—are remarkably similar across marriage types, suggesting no clear descriptive differences in reported victimization between women in consanguineous and non-kin marriages. This finding is broadly consistent with evidence from some Muslim-majority settings indicating that women in consanguineous marriages may experience lower levels of spousal violence than those in non-consanguineous unions (Bhatta and Haque, 2015; Hamamy and Alwan, 2016). However, the evidence remains mixed. For example, Rahaman et al. (2022) report higher rates of spousal violence among women in consanguineous marriages in India. Differences across studies may reflect variations in institutional and cultural contexts, sample composition, or differences in women's willingness to disclose experiences of violence.[6]

**Independent Variables:**

*Consanguineous marriage:* The PDHS 2017–18 provides detailed information on whether a woman is related to her husband by blood and, if so, the specific nature of that relationship. Using this information, we construct a variable for consanguineous marriage that equals 1 if the respondent reports any blood relationship with her husband (including first cousins, second cousins, or other relatives), and 0 if the marriage is outside the kin group.

*Control variables:* To capture the degree of kinship consolidation within marriage, we employ two complementary proxy measures. First, we use the number of household members as an indicator of residence within an extended family system. In the Pakistani context, larger household size is strongly associated with multigenerational and joint family living arrangements, which often coincide with intensified kin oversight, shared economic resources, and collective enforcement of gender norms. Such extended family systems can amplify patriarchal authority by embedding women within dense kinship networks that limit privacy, autonomy, and exit options. Second, we use women's reported involvement in choosing their husbands as a proxy for the degree of marital consolidation within kin networks. In contexts of consanguineous marriage, a woman's choice to marry a kin may reflect strategic conformity, in which she may be pressured or strategically motivated to choose blood relatives as spouses to preserve family ties, economic security, and social legitimacy.

[6] Accordingly, the descriptive statistics should not be interpreted as evidence that consanguineous marriage has no relationship with women's exposure to violence, but rather that no substantial differences in reported victimization are evident in the PDHS sample.

Drawing on prior research (Akbary et al., 2022; Nasrullah et al., 2017; Jesmin, 2017; Ahmad et al., 2021; Saud et al., 2021; Pettitt et al., 2024; Shreemoyee et al., 2025), we also include several independent variables to control for individual, household, and regional characteristics. These controls include women's age and education (years of schooling), which capture life-cycle and human capital effects. Older women have been found to exhibit greater acceptance of IPV in some settings (e.g., Sayem et al., 2012), while education has consistently been associated with more egalitarian gender norms and lower acceptance of IPV (Sardinha and Nájera Catalán, 2018). Husband's education (years of schooling) is included to account for household socioeconomic position, while household wealth quintiles help control for economic security and material deprivation. We also control for residence in urban areas (captured through a dummy variable, where 1 = resides in an urban area; 0 = resides in a rural area) to capture differential exposure to social norms and institutional environments.

**Table 1: Descriptive Statistics**

| Variable | Non-Kin Marriage | Consanguineous Marriage | Total |
|---|---|---|---|
| **Dependent variables: Women's acceptance of IPV (1 = justified; 0= IPV is not justified)** | | | |
| Going out without permission | 0.323 (0.468) | 0.399 (0.490) | 0.370 (0.483) |
| Neglecting children | 0.269 (0.444) | 0.330 (0.470) | 0.307 (0.461) |
| Arguing with husband | 0.332 (0.471) | 0.388 (0.487) | 0.366 (0.482) |
| Refusing sexual intercourse | 0.279 (0.449) | 0.338 (0.473) | 0.316 (0.465) |
| Burning food | 0.171 (0.376) | 0.209 (0.406) | 0.194 (0.396) |
| **Control variables (continuous)** | | | |
| Age (years) | 32.45 (8.07) | 32.03 (8.37) | 32.19 (8.26) |
| Education (years) | 5.38 (5.58) | 4.03 (4.99) | 4.54 (5.26) |
| Husband's education (years) | 7.45 (5.34) | 6.83 (5.21) | 7.07 (5.27) |
| Number of household members | 8.02 (4.32) | 8.68 (4.69) | 8.43 (4.57) |
| **Binary controls (proportions)** | | | |
| Say in choosing husband (1=had a say in choosing husband; 0=didn't have a say in choosing husband) | 0.771 | 0.829 | 0.807 |
| Urban (1=resides in urban area; 0= resides in rural area) | 0.532 | 0.454 | 0.484 |
| **Past experience with violence (proportions)** | | | |
| Experienced any less severe violence | 0.249 (0.43) | 0.244 (0.43) | 0.244 (0.43) |

| Variable | Non-Kin Marriage | Consanguineous Marriage | Total |
|---|---|---|---|
| Experienced any severe violence | 0.070 (0.255) | 0.067 (0.251) | 0.068(0.253) |
| Experienced any sexual violence | 0.049 (0.217) | 0.042 (0.201) | 0.045(0.207) |
| **Wealth index (%)** | | | |
| Poorest | 15.15 | 21.58 | 19.15 |
| Poorer | 19.53 | 22.69 | 21.50 |
| Middle | 19.18 | 19.99 | 19.68 |
| Richer | 21.09 | 17.90 | 19.10 |
| Richest | 25.05 | 17.84 | 20.56 |
| **Region (%)** | | | |
| Punjab | 22.85 | 22.39 | 22.56 |
| Sindh | 14.66 | 20.31 | 18.17 |
| KPK | 16.61 | 15.28 | 15.78 |
| Balochistan | 8.78 | 13.06 | 11.44 |
| Gilgit-Baltistan | 9.38 | 4.81 | 6.53 |
| Other | 27.73 | 24.17 | 25.51 |
| **Sample size** | 5,684 | 9,381 | 15,065 |

Notes: Standard deviations in parentheses for continuous variables. For binary variables, table entries report proportions. For categorical variables (wealth index and region), column percentages are reported.

Previous studies also indicate that women's past experience with violence can influence their attitudes toward their acceptance of IPV (Sayem et al., 2012). Therefore, controlling for this factor is necessary. We include three control variables that represent experiences of less severe, severe, and sexual violence.[7] Finally, we include region-fixed effects to account for deep-rooted cultural, linguistic, and historical differences in kinship practices and gender norms across Pakistan.

**Dependent Variable:**

*Acceptance of IPV:* The dependent variable is women's acceptance of IPV in various scenarios related to traditional gender roles. This variable is derived from a set of attitudinal questions from the Pakistan Demographic and Health Survey, which assess normative approval of spousal

[7] Pakistan Demographic and Health Survey (PDHS) 2017–18 captures these variables as follows. Any less severe physical violence (recorded as d106 in PDHS data): a composite indicator coded 1 if the respondent has ever been pushed, shaken, or had something thrown at her (d105a); slapped (d105b); punched with a fist or hit with something harmful (d105c); or had her arm twisted or hair pulled (d105j). Any severe physical violence (d107): a composite coded 1 if she has ever been kicked or dragged (d105d); strangled or burnt (d105e); or threatened with a knife, gun, or other weapon (d105f). Any sexual violence (d108): a composite coded 1 if she has ever been physically forced into unwanted sex (d105h); forced into other unwanted sexual acts (d105i); or physically forced to perform sexual acts she did not want to (d105k).

violence—referred to in the survey as "wife-beating"—under five common scenarios: if a wife goes out without informing her husband, neglects the children, argues with her husband, refuses sexual intercourse, or burns the food. Each question is coded as a binary indicator, with one indicating that the respondent believes a husband is justified in hitting or beating his wife in that situation, and zero if not. Consistent with existing research, these measures reflect constrained agency and strategic conformity within patriarchal social structures rather than endorsing violence itself (Kandiyoti, 1988; Yount et al., 2013; Tayyab et al., 2017; Akbary et al., 2022).

Table A1 in the appendix presents an analysis of IPV acceptance among Pakistani men and women. The data reveal a consistent and significant gender gap across all five scenarios, with women consistently reporting higher acceptance of IPV than men. The most common scenarios are leaving the house without permission and arguing with the husband, with approximately 35% of women justifying IPV in these cases, compared to 27% and 20% of men, respectively. The disparity increases sharply for refusing sexual intercourse, where 29.5% of women versus only 9.6% of men see violence as justified, and for neglecting children, with nearly 29% of women and 18% of men. Burning food is the least accepted justification of IPV from both genders, but the gap remains large: 18% of women versus nearly 4% of men justify IPV in this context. In all scenarios, women are two to three times more likely than men to justify IPV. While this pattern may seem surprising initially, it aligns with research indicating that women in patriarchal societies often internalize gender norms, especially among ever-married women whose socialization reinforces deference.

There is also a striking regional variation in women's acceptance of IPV in Pakistan (see Table A2 in the Appendix), reflecting deep heterogeneity in gender norms, regional institutional structures, and customary practices.

The Federally Administered Tribal Areas (FATA) —which merged intoKhyber Pakhtunkhwa (KPK) in 2018—have the highest acceptance of IPV[8]. Acceptance rates across all five scenarios are dramatically higher than any other region, ranging from 43% for burning food to 89% for both going out without the husband's permission and for arguing with the husband. It is important to note that FATA was governed separately under the Frontier Crimes Regulations, which excluded the population from Pakistan's civil law and legal protections for decades. The region is characterized by deeply entrenched tribal customary codes that grant near-absolute authority to male household heads.

In other situations, KPK, Pakistan's northwestern province bordering Afghanistan, and Balochistan, the southwestern province bordering both Afghanistan and Iran, demonstrate the next-highest levels of acceptance. However, these are still considerably below the levels seen in FATA. Acceptance of IPV ranges from about 23% for burning food to 55% for arguing in KPK, while in Balochistan it ranges from 22% to 56%. Both areas share traits such as largely rural populations, strong patrilineal tribal systems, lower female educational attainment and labor force participation, and a relatively limited reach of state institutions. These factors are consistently linked in the literature to higher acceptance of IPV (Yount et al., 2013; Jesmin, 2017; Sardinha and Nájera Catalán, 2018). Gilgit-Baltistan shows a regionally similar pattern[9], with acceptance

---

[8] These regional patterns in women's acceptance of IPV align with those observed in men. In the men's sample, consisting of a separate sub-sample of 3,691 ever-married men aged 15–49, similar regional patterns are also evident.
[9] Gilgit-Baltistan (GB), directly bordering KPK province of Pakistan, is an administrative jurisdiction which remains under constitutional ambiguity. To contextualize the distinct socioeconomic and institutional conditions, it's important note that while GB is governed by Pakistan administratively but remains without the legal protections or

rates between 30% and 47% depending on the scenario, somewhat lower than in KPK and Balochistan despite its geographic remoteness. This may partly reflect the Aga Khan Development Network's long-standing presence in the region, which has invested heavily in female education and community development.[10]

Punjab and Sindh cluster at moderate levels, with acceptance rates between 22–28% for the most supported scenarios, dropping to about 16–18% for burning food. Punjab is the most economically developed and urbanized province in Pakistan, with higher female educational attainment[11] and distinct social and cultural practices. The ICT region, which includes Islamabad, Pakistan's capital, has the lowest acceptance rates for IPV, with rates as low as 6% for burning food and 15% for going out without permission. This region is highly urbanized, with high educational attainment and institutional protections.

The scenario-specific pattern is consistent across all regions: burning food invariably has the lowest acceptance, whereas going out without the husband's permission or arguing with the husband has the highest acceptance, suggesting that the underlying normative hierarchy of justifications is broadly shared across regions, even as the absolute level of acceptance varies enormously. Regional variation in Pakistan is not marginal—the gap between FATA and ICT is roughly 70 percentage points across several scenarios, underscoring that any analysis of IPV acceptance in Pakistan that does not account for regional fixed effects risks severely misattributing variation to individual-level factors.

Understanding the regional, institutional, and cultural context is essential, as research shows that high acceptance of IPV is positively correlated with rural residence (Pettitt et al., 2024; Gunarathne et al., 2026), lower educational attainment (Tran et al., 2016; Biswas et al., 2017; Shreemoyee et al., 2025), unclear legal protections for women (Shams et al., 2020), and entrenched customary practices and discriminatory social institutions (Sardinha and Nájera Catalán, 2018; Nadeem and Malik, 2021). Tran et al. (2016) found that over 90% of women in Afghanistan justify IPV in certain situations. The study's cross-country analysis also shows that, at the country level, acceptance of IPV is closely linked to the Human Development Index (HDI), the Gender Inequality Index, and average years of schooling. Countries with lower development, greater gender inequality, and lower educational attainment tend to have higher IPV acceptance among both men and women. At the individual level, factors such as rural residence, belonging to the poorest wealth quintile, lower education levels, younger age, and marriage history are independently associated with greater acceptance among women.

## V. Results and Discussion

### Examining the Intensity of Women's Attitudes Toward IPV

---

representation. See Dawn Editorial Board. (2025, January 21). GB's status. *Dawn*. https://www.dawn.com/news/1886684

[10] See https://the.akdn/en/where-we-work/south-asia/pakistan/conservation-and-development-in-gilgit-baltistan-pakistan

[11] A recent Gallup Pakistan report analyzed the Provincial Census Report Punjab 2017 from the Pakistan Bureau of Statistics and found that Punjab's overall literacy rate is 64%, roughly five percentage points above Pakistan's national average of 59%. Comparing the 1998 and 2017 censuses, female literacy improved by about 22 percentage points over that period, outpacing the male gain of roughly 14 points (Gallup Pakistan, 2024).

Before exploring the role of consanguineous marriages in attitudes towards IPV, we examine the intensity of these attitudes in the sample. To move beyond binary measures of IPV acceptance, we construct an IPV index that captures the intensity of women's acceptance of IPV across the five standard DHS scenarios. The index is constructed by summing each woman's binary responses to the five wife-beating justification questions, yielding a score that ranges from 0 to 5. A value of 0 indicates that a woman rejects IPV in every scenario, while a value of 5 indicates that she considers violence justified in all five situations. Unlike the scenario-specific binary indicators used in the main analysis, the IPV index treats acceptance as a matter of degree rather than a categorical yes-or-no judgment, allowing for an examination of how broadly and deeply women internalize norms that legitimize IPV.

Table 2 displays how the IPV index is spread across the entire sample. The distribution is notably polarized: over half of the women (54.4%) have an IPV index of zero, showing complete rejection of IPV in all five scenarios. Conversely, 14.4% score five, meaning they justify violence in all five scenarios. The middle categories are less common, with 6.4% justifying IPV in exactly one scenario. The remaining 23.8% fall between 2 and 4, indicating partial but significant acceptance of IPV norms. Overall, 45.6% of women justify IPV in at least one situation. This pattern, with women predominantly at either extreme of rejection or acceptance and fewer in the middle, indicates that attitudes toward IPV in Pakistan are not smoothly distributed but reflect distinct orientations about the legitimacy of IPV, which we explore further through the lens of marriage dynamics.

**Table 2: Intensity of Women's Acceptance of IPV: IPV Index Distribution**

| IPV index | Description | Frequency | Percent |
|---|---|---|---|
| 0 | Rejects IPV in all five scenarios | 8197 | 54.41 |
| 1 | Justifies IPV in one scenario | 966 | 6.41 |
| 2 | Justifies IPV in two scenarios | 1297 | 8.61 |
| 3 | Justifies IPV in three scenarios | 1334 | 8.85 |
| 4 | Justifies IPV in 4 scenarios | 1098 | 7.29 |
| 5 | Justifies IPV in all five scenarios | 2173 | 14.42 |

Notes: The IPV index is constructed by summing women's responses to the five DHS questions on whether a husband is justified in beating his wife if she (i) goes out without telling him, (ii) neglects the children, (iii) argues with him, (iv) refuses to have sexual intercourse, or (v) burns the food. Each affirmative response is coded as 1 and summed to produce an index ranging from 0 to 5. Higher values indicate acceptance of IPV in a greater number of scenarios, with 0 representing rejection of IPV in all five situations and 5 representing justification of IPV in every situation.

## Consanguineous Marriage and Women's Acceptance of IPV

Our analysis in Table 3 expands to compare the distribution of the IPV index between women in consanguineous and non-kin marriages. Several notable patterns emerge. First, women in consanguineous marriages are significantly less likely to fully reject IPV. Specifically, 59.4 percent of women in non-kin marriages score zero on the IPV index, indicating rejection across all five scenarios. This percentage drops to 51.4 percent among women in consanguineous marriages, a difference of eight points. Second, women in consanguineous marriages are more likely to justify IPV in all five scenarios. Among non-kin married women, 12.7 percent score five, compared to 15.5 percent of those in consanguineous marriages—a nearly three-point difference. Third, women in consanguineous marriages also dominate the intermediate categories (levels 1-4). While 27.9 percent of non-kin married women fall in these categories, 33.0 percent of consanguineous married women do. The differences at levels 2, 3, and 4 are statistically significant at the 1 percent level, whereas the difference at level 1 is not, indicating that persistent acceptance of IPV across multiple scenarios, rather than situational or marginal acceptance, is associated with marriage type.

**Table 3: Women's Acceptance of IPV by Marriage Type**

| IPV Acceptance Index | Non-Kin Marriage (N, %) | Consanguineous Marriage (N, %) | Total (N) | Statistical significance |
|---|---|---|---|---|
| 0 (Rejects IPV in all 5 scenarios) | 3,377 (59.4%) | 4,820 (51.4) | 8,197 | Sign. |
| 1 (Lowest acceptance) | 362 (6.4) | 604 (6.4) | 966 | Not. Sign. |
| 2 | 411 (7.2) | 886 (9.4) | 1,297 | Sign. |
| 3 | 439 (7.7) | 895 (9.5) | 1,334 | Sign. |
| 4 | 373 (6.6) | 725 (7.7) | 1,098 | Sign. |
| 5 (Highest acceptance) | 722 (12.7) | 1,451 (15.5) | 2,173 | Sign. |
| Total | 5,684 (38) | 9,381 (62%) | 15,065 | |

Notes: The IPV Acceptance Index ranges from 0 to 5 and captures the intensity of women's acceptance of intimate partner violence (IPV) across five scenarios included in the Pakistan Demographic and Health Survey (PDHS) 2017-18. Respondents were asked whether a husband is justified in hitting his wife if she: goes out without informing him, neglects the children, argues with him, refuses to have sexual intercourse with him, or burns the food. Here, the IPV index captures the intensity of these attitudes. A value of 0 indicates that the respondent rejects IPV in all five scenarios. Values 1–4 indicate that the respondent justifies IPV in one to four scenarios. A value of 5 represents the highest level of acceptance, where the respondent justifies IPV in all five scenarios. Chi-squared ( $\chi^2$) tests for differences in proportions are conducted for statistical significance. Sign. means the difference is statistically significant at the 1% level.

Table 4 presents logistic regression results on the association between consanguineous marriage and women's acceptance of IPV across five scenarios, while accounting for individual, household, and regional factors. Consanguineous marriage significantly increases the likelihood of justifying IPV in four of five scenarios: going out without permission (OR = 1.52), refusing sexual intercourse (OR = 1.50), neglecting children (OR = 1.34), and arguing with the husband (OR = 1.20). The exception is burning food, where the effect is minimal and not statistically significant.

**Table 4: Consanguinity and Women's Acceptance of IPV: Logistic Regression Results**

| Variable | (1) Going out OR [95% CI] | (2) Neglecting children OR [95% CI] | (3) Arguing OR [95% CI] | (4) Refusing intercourse OR [95% CI] | (5) Burning food OR [95% CI] |
|---|---|---|---|---|---|
| Consanguineous marriage | **1.523***** [1.291, 1.796] | **1.340***** [1.139, 1.576] | **1.203**** [1.020, 1.417] | **1.498***** [1.267, 1.772] | 1.056 [0.876, 1.272] |
| Age | 0.993 [0.984, 1.002] | 0.997 [0.988, 1.007] | 1.002 [0.993, 1.012] | 1.000 [0.990, 1.010] | 0.998 [0.987, 1.008] |
| Respondent's education (years) | **0.939***** [0.920, 0.960] | **0.930***** [0.910, 0.950] | **0.930***** [0.910, 0.950] | **0.938***** [0.918, 0.959] | **0.918***** [0.893, 0.944] |
| Husband's education (years) | 0.994 [0.976, 1.012] | 0.988 [0.971, 1.006] | 1.003 [0.985, 1.021] | 1.009 [0.991, 1.028] | 0.986 [0.966, 1.006] |
| Wife's had a say in choosing husband | 1.052 [0.870, 1.272] | **1.439***** [1.188, 1.743] | 0.982 [0.813, 1.187] | **0.772***** [0.640, 0.933] | 1.159 [0.937, 1.434] |
| **Wealth index (ref: poorest)** | | | | | |
| Poorer | **0.614***** [0.488, 0.772] | **0.704***** [0.567, 0.875] | **0.571***** [0.454, 0.719] | **0.625***** [0.499, 0.783] | **0.624***** [0.495, 0.786] |
| Middle | **0.402***** [0.309, 0.524] | **0.510***** [0.394, 0.661] | **0.420***** [0.322, 0.548] | **0.434***** [0.333, 0.565] | **0.421***** [0.315, 0.563] |
| Richer | **0.412***** [0.304, 0.558] | **0.479***** [0.355, 0.647] | **0.390***** [0.288, 0.529] | **0.388***** [0.285, 0.528] | **0.395***** [0.278, 0.559] |
| Richest | **0.252***** [0.174, 0.364] | **0.274***** [0.189, 0.399] | **0.207***** [0.143, 0.301] | **0.199***** [0.135, 0.293] | **0.160***** [0.097, 0.266] |
| Number of household members | 1.017 [0.995, 1.039] | 1.016 [0.995, 1.038] | **1.022*** [1.000, 1.044] | **1.020*** [0.998, 1.043] | 1.016 [0.992, 1.042] |
| Urban residence (ref: rural residence) | 0.924 [0.770, 1.108] | 1.016 [0.847, 1.219] | 0.994 [0.829, 1.192] | 0.924 [0.768, 1.111] | 0.883 [0.714, 1.093] |
| **Past experience with violence** | | | | | |
| Experienced any less severe violence | **1.296**** [1.054, 1.595] | 0.871 [0.709, 1.071] | **1.264**** [1.027, 1.556] | **1.219*** [0.992, 1.498] | **0.729***** [0.574, 0.926] |
| Experienced any severe violence | **1.431**** [1.019, 2.010] | 0.970 [0.689, 1.367] | 1.248 [0.891, 1.749] | **1.416**** [1.009, 1.989] | 0.965 [0.649, 1.434] |
| Experienced any sexual violence | **1.444*** [0.985, 2.118] | **1.959***** [1.350, 2.841] | 1.302 [0.883, 1.920] | **1.397*** [0.957, 2.041] | **1.672**** [1.109, 2.520] |
| **Region fixed effects (ref: Punjab)** | | | | | |
| Sindh | **0.764**** [0.592, 0.986] | 0.856 [0.665, 1.103] | **0.644***** [0.498, 0.832] | **0.467***** [0.355, 0.614] | **0.724**** [0.538, 0.973] |
| KPK | **3.055***** [2.397, 3.892] | **2.183***** [1.711, 2.785] | **3.346***** [2.630, 4.259] | **2.259***** [1.770, 2.884] | **1.274*** [0.956, 1.698] |
| Balochistan | **2.041***** [1.541, 2.704] | 1.047 [0.783, 1.400] | **1.324**** [1.001, 1.752] | **1.280*** [0.962, 1.705] | **0.743*** [0.527, 1.047] |
| Gilgit-Baltistan | **2.337***** [1.700, 3.213] | **1.880***** [1.365, 2.589] | **1.457**** [1.057, 2.010] | **1.640***** [1.189, 2.262] | **1.403*** [0.981, 2.007]) |
| ICT[12] | 0.783<br>[0.526, 1.166] | **0.638****<br>[0.417, 0.976] | 0.916<br>[0.628, 1.336] | 0.751<br>[0.498, 1.133] | 0.650<br>[0.376, 1.125] |
| AJK[13] | 0.862<br>[0.649, 1.146] | 0.795<br>[0.595, 1.063] | **0.784***<br>[0.591, 1.041] | **0.662*****<br>[0.492, 0.890] | 0.805<br>[0.573, 1.131] |
| FATA[14] | **15.430*****<br>[9.781, 24.342] | **3.170*****<br>[2.286, 4.396] | **19.185*****<br>[11.510, 31.977] | **4.226*****<br>[2.987, 5.980] | **2.429*****<br>[1.726, 3.418] |
| Observations | 3,864 | 3,856 | 3,861 | 3,817 | 3,866 |
| Pseudo $R^2$ | 0.2086 | 0.1313 | 0.2080 | 0.1761 | 0.1373 |

**Notes:** Dependent variables are binary indicators of a wife's acceptance of intimate partner violence. Each variable is coded 1 if a woman believes a husband is justified in hitting his wife for the following reason, and 0 otherwise: leaving the house without informing him (*going out*), neglecting the children (*neglecting children*), arguing with him (*arguing*), refusing sexual intercourse (*refusing intercourse*), or burning the food (*burning food*). The table reports odds ratios along with 95% confidence intervals in brackets. All models were estimated on the domestic violence subsample of the PDHS 2017–18. * $p < 0.10$ ** $p < 0.05$ *** $p < 0.01$

In Table 4, women's education consistently serves as a protective factor, lowering the odds of IPV acceptance in all five scenarios. This aligns with existing research suggesting that higher education improves awareness of personal rights and protections (Shams et al., 2020; Pettitt et al., 2024). Husband's education, however, is not statistically significant in any scenario. Consistent with prior studies (see Shreemoyee et al., 2025), household wealth shows a strong, monotonic gradient—women in the richest quintile are less likely to justify IPV than those in the lowest quintile.

Household size shows small yet marginally significant positive associations with acceptance of arguing (OR = 1.022) and refusal of intercourse (OR = 1.020), suggesting that living in extended families may increase normative surveillance and enforcement of gendered behavioral norms. However, urban versus rural residence is not statistically significant in any model, indicating that, after controlling for wealth, education, and regional effects, urban residence alone does not independently predict lower IPV acceptance. A wife's say in choosing a husband increases acceptance of IPV in the scenario involving neglecting children, but decreases acceptance in the scenario involving refusal of intercourse. This mixed pattern suggests that marital agency has scenario-specific effects rather than providing overall protection.

Women who have experienced less severe physical violence are significantly more likely to justify IPV when going out (OR = 1.30), arguing (OR = 1.26), and marginally more likely to justify refusing intercourse (OR = 1.22). They are less likely to justify burning food (OR = 0.73). Women who experienced severe physical violence also show higher odds of justifying going out (OR = 1.43) and refusing intercourse (OR = 1.42).

[12] Islamabad Capital Territory (ICT) is the federal jurisdiction that includes Islamabad, the capital city of Pakistan. It is a small, densely urbanized area managed directly by the federal government.

[13] AJK (Azad Jammu and Kashmir) is a special territory administered by Pakistan, located in the northwestern region of the country.

[14] FATA (Federally Administered Tribal Areas) was a former administrative designation for a belt of tribal territories along Pakistan's northwestern border with Afghanistan. FATA was historically governed under a separate legal framework (the Frontier Crimes Regulations), which gave the federal government direct authority and excluded residents from many protections under Pakistani civil and criminal law. In May 2018, FATA was formally merged into Khyber Pakhtunkhwa (KPK) province, ending its semi-autonomous status. However, by the time PDHS 2017–18 was conducted, the merger was underway, and FATA was reported separately.

We also find that sexual violence correlates strongly with acceptance of neglecting children (OR = 1.96) and burning food (OR = 1.67). These findings support the normalization hypothesis—that exposure to violence can adjust women's perceptions of acceptable spousal behavior—though the different domains suggest that the relationship varies and may involve different mechanisms depending on the type of violence experienced. This result is consistent with the related literature that women who had personally experienced IPV in the past year were notably more likely to justify it, which aligns with the theoretical model of high social justification environments normalizing violence for victims themselves (Waltermaurer, 2012).

Consistent with the descriptive analysis, regional effects are large, with the FATA region exhibiting extremely high odds of IPV acceptance (up to OR = 19 for arguing). Additionally, KPK and Balochistan have significantly higher odds than Punjab. This suggests that women's acceptance of IPV in Pakistan is not a uniform cultural phenomenon but is deeply embedded in region-specific institutional, historical, and normative contexts. Where a woman lives—and the governance structures, customary codes, and institutional arrangements that characterize her region—shapes her attitudes toward IPV at least as much as her own socioeconomic characteristics.

**Women's Empowerment in Household Decision-Making**

To assess women's decision-making power within households, we create two binary measures of empowerment based on standard DHS autonomy variables. These variables indicate who typically makes decisions about the respondent's healthcare (v743a), major household purchases (v743b), visits to family or relatives (v743d), handling the husband's earnings (v743f), and spending the respondent's income (v739). For each domain, a woman is coded 1 if she reports making decisions alone or jointly with her husband, and 0 if decisions are made by her husband or another. The first measure, *empowerment (all domains)*, is 1 if a woman participates in decision-making across all five domains, indicating full household empowerment, and 0 otherwise. The second measure, *empowerment (any domain)*, is 1 if a woman participates in at least one domain, indicating some degree of household autonomy, and 0 if she has no decision-making role in any domain. These two measures help distinguish between full empowerment and partial participation, highlighting the variation in women's agency observed in Pakistan (Lassi et al., 2021).

Using similar measures, Malik et al. (2024) found that women's empowerment in the household serves as a protective force, lowering women's acceptance of IPV in Pakistan. This analysis therefore examines whether including women's household decision-making (a proxy for women's empowerment) changes the main finding that consanguineous marriage correlates with greater acceptance of IPV. These models also assess whether women's household empowerment predicts lower acceptance across all five scenarios. The results are presented in tables A3-A5 in the Appendix. Three important findings stand out. First, the most important finding across all five models in the Appendix is that the consanguinity coefficient is remarkably stable after controlling for empowerment. For going out without permission, the odds ratio remains at 1.52–1.54 regardless of which empowerment measure is included. For arguing, it holds at approximately 1.20 in both specifications. For refusing sexual intercourse and neglecting children, it remains significant at 1.51 and 1.35, respectively. For burning food, it remains small and insignificant in all specifications. This stability confirms that the association between consanguineous marriage and IPV acceptance is not simply a proxy for women's lack of household decision-making power. Consanguinity appears to operate through a distinct channel—most likely the normative and surveillance dimensions of overlapping kinship networks—that is not captured by individual-level empowerment measures.

Second, the comparison between Model 1 and Model 2 in Tables A3 and A4 highlights an important distinction. Women's empowerment, when defined as participation in all five decision-making areas, is not statistically significant in any of the five IPV scenarios (OR = 0.938 for going out; OR = 0.762 for arguing). This indicates that complete household empowerment is rare in the sample and does not independently predict IPV acceptance after controlling for other factors. Conversely, empowerment, defined as participation in at least one decision-making domain, is consistently significant and protective across all five scenarios, reducing the odds of accepting IPV by a substantial amount, depending on the scenario. This indicates that even partial involvement in decision-making may influence women's agency and attitudes, and the level of empowerment needed to affect normative acceptance is relatively low. From a social policy perspective, this suggests that small improvements in women's household agency—not necessarily full empowerment—may substantially reduce women's acceptance of IPV.

Finally, consistent with the literature, women's empowerment serves as a protective factor and is statistically significant across all five IPV scenarios. However, the magnitude of this effect varies. The strongest protective effect appears for burning food (OR = 0.622), followed by going out (OR = 0.645), refusing sexual intercourse (OR = 0.702), arguing (OR = 0.712), and neglecting children (OR = 0.733). The finding that empowerment appears to reduce women's acceptance even for burning food—the scenario where consanguinity showed no effect under other models—suggests that women's decision-making power operates more broadly across all IPV domains, whereas consanguinity's effect is more selective, concentrated in areas directly related to women's bodily autonomy. In other words, empowerment and consanguinity appear to be distinct and largely independent pathways. Consanguinity seems to selectively reinforce patriarchal control over women's mobility and bodily autonomy, while low household decision-making power broadly increases justification of violence across all scenarios.

Results from the ordered logistic regression in Table 5 indicate that consanguineous marriage is significantly associated with a higher intensity of accepting intimate partner violence. Women in consanguineous marriages show greater odds of accepting IPV across multiple scenarios, even after controlling for women's and husbands' education, household wealth, residence, ethnicity, and marital choice. This finding aligns with the argument that consanguineous marriage operates as an institutional mechanism reinforcing patriarchal norms that legitimize violence as discipline.

**Table 5: Consanguinity and Women's Acceptance of IPV: Ordered Logistic Regression Results**

| Variables | IPV Acceptance Index |
|---|---|
| Consanguineous marriage (1 = married to kin) | **0.106***** (0.035) |
| Age | −0.000 (0.002) |
| Respondent's education (years) | **−0.068***** (0.004) |
| Husband's education (years) | −0.001 |

| Variables | IPV Acceptance Index |
|---|---|
| | (0.004) |
| Wife had a say in choosing husband | **0.146***** (0.042) |
| Household wealth (ref: poorest) | |
| Poorer | **−0.292***** (0.049) |
| Middle | **−0.629***** (0.056) |
| Richer | **−0.839***** (0.065) |
| Richest | **−1.22***** (0.076) |
| Urban residence | **−0.140***** (0.039) |
| No. of household members | **0.027***** (0.003) |
| Cut point 1[15] | 0.078 |
| Cut point 2 | 0.403 |
| Cut point 3 | 0.867 |
| Cut point 4 | 1.407 |
| Cut point 5 | 1.958 |
| Observations | 14,420 |
| Log likelihood | −18,807 |
| LR $\chi^2$ (16) | 3032.8 |
| Pseudo $R^2$ | 0.074 |

Notes: The table reports coefficients from an ordered logistic regression estimating women's acceptance of intimate partner violence (IPV). The dependent variable is an ordinal IPV acceptance index ranging from 0 (does not justify IPV in any scenario) to 5 (justifies IPV in all five scenarios). Positive coefficients indicate higher odds of endorsing IPV across more scenarios. Robust standard errors are reported in parentheses. Models control for region-fixed effects and past experience with violence. *** $p < 0.01$, ** $p < 0.05$, * $p < 0.10$.

Consistent with the previous results, women's education and household wealth emerge as strong protective factors, consistently associated with reduced IPV acceptance, in line with theories emphasizing fallback positions and economic security. Similarly, residence in urban areas is associated with a lower acceptance of IPV. Notably, women's involvement in choosing their husbands appears positively associated with IPV acceptance. This may suggest that women who actively choose their relatives as their spouses tend to engage in strategic conformity within consolidated kinship systems. This interpretation is reinforced by the positive association between household size and acceptance of IPV, suggesting that extended-family living arrangements intensify normative surveillance and the enforcement of gendered discipline.

[15] In the ordered logistic regression, the estimated cut points represent threshold parameters that map the unobserved latent index of attitudes toward intimate partner violence (IPV*) onto the observed ordered outcome categories. Specifically, cut1 denotes the threshold separating women who do not justify IPV in any scenario (IPV index = 0) from those who justify IPV in at least one scenario (≥1). Cut2 separates justification in at most one scenario from justification in two or more scenarios, cut3 distinguishes up to two versus three or more scenarios, cut4 distinguishes up to three versus four or more scenarios, and cut5 separates justification in up to four scenarios from justification in all five scenarios.

**Results Under Propensity Score Matching**

Women in consanguineous marriages differ from women in non-kin marriages in many ways beyond marriage type itself, such as education, household wealth, household size, place of residence, or region. If we simply compared the two groups directly, these background differences could confound the relationship between consanguinity and attitudes toward IPV.

Propensity score matching (PSM) helps address this problem by making the two groups as similar as possible before comparing their attitudes. Rosenbaum and Rubin (1983) defined the propensity score as the conditional probability of assignment to a particular treatment given the observed covariates. The propensity score captures the probability (bounded between 0 and 1) for an individual to be exposed to a treatment or risk factor based on their observed baseline characteristics. In observational research, the primary purpose of the propensity score is to adjust for measured confounding by balancing observed characteristics between treatment and control groups. Thus, propensity score methods attempt to account for systematic differences in baseline covariates, mimicking the conditions of random assignment that would have been achieved through randomized controlled trials (Chesnaye et al., 2022).

The method is applied in three steps. First, each woman is assigned a propensity score, which represents her predicted likelihood of being in a consanguineous marriage based on observable characteristics such as age, education, husband's education, household wealth, number of household members, rural–urban residence, and other control variables reported above. Women with similar characteristics therefore receive similar scores, regardless of whether they are actually in a consanguineous marriage. Second, each woman in a consanguineous marriage is matched to a woman in a non-kin marriage who has a very similar propensity score. This creates pairs of women who are comparable in terms of observed background characteristics but differ in whether they married a relative. Finally, the analysis compares attitudes toward IPV only within these matched pairs. Because the matched women are similar on observable traits, differences in acceptance of IPV between them are less likely to reflect socioeconomic or demographic differences and more likely to reflect differences in marriage type itself.

Nearest-neighbor propensity score matching is one of the most commonly used methods to estimate treatment effects in observational studies, pairing treated and control units with similar propensity scores to balance observed covariates before outcome comparison (Stuart, 2010). Caliper matching—which restricts matches to pairs within a specified maximum difference in propensity scores—is widely recommended to improve the quality of matches and reduce residual bias (Austin, 2010; Stuart, 2010; Lunt, 2014). The use of a common-support restriction, which excludes treated or control units with no comparable matches, further improves the credibility of causal estimates (Rosenbaum and Rubin, 1983).

To address potential selection bias arising from systematic differences between women in consanguineous and non-kin marriages, we use PSM to estimate the effect of consanguineous marriage on women's acceptance of IPV across five scenarios: going out without permission, neglecting children, arguing with husband, refusing sexual intercourse, and burning food. We estimate propensity scores using probit regression, modeling the likelihood of being in a consanguineous marriage as a function of respondent age, education, husband's education, involvement in choosing husband, household wealth quintiles, number of household members, urban or rural residence, as well as controls for region. We implement nearest-neighbor matching with a 1:1 ratio and imposed a highly conservative caliper of 0.05, restricting matches to pairs with propensity score differences of no more than five percentage points, nearly four to five times more restrictive than standard practice (Stuart, 2010; Austin, 2010, 2014). We enforce the common support restriction, dropping observations outside the region of overlapping propensity score distributions (7-–8 observations per model, representing <0.06% of the sample). This matching strategy achieves excellent covariate balance, reducing mean standardized bias from 14.5–14.6% before matching to 1.2–2.2% after matching, with pseudo $R^2$ values falling from 0.023 to 0.000-0.001, indicating that observables have virtually no remaining predictive power for treatment assignment in the matched sample. We calculate the average treatment effect on the treated (ATT) for each outcome—the percentage-point difference in IPV acceptance between women in consanguineous marriages and their matched non-consanguineous counterparts.

Table 6 presents propensity score matching estimates of the association between consanguineous marriage and women's acceptance of IPV across five scenarios. In four out of five cases, consanguineous marriage is significantly associated with a higher likelihood that women justify IPV, even after accounting for socioeconomic, demographic, household, and regional factors. Women in consanguineous marriages are 4.1 percentage points more likely to justify IPV if they go out without informing their husband ($p < 0.01$), 2.4 points more likely if they neglect the children ($p < 0.05$), and 2.3 points more if they argue with their spouse ($p < 0.10$). Overall, the strongest link is observed for going out without husband's permission ( 4.1 percentage points, $p < 0.01$) followed by refusal of sexual intercourse (3 percentage points, $p < 0.01$). In contrast, the effect is smaller for the scenario where a woman burns food (1.5 percentage points) and not statistically significant ($p > 0.1$). Overall, the findings are consistent with the theoretical framework that consanguineous marriages reinforce patriarchal control, particularly by shaping women's agency and autonomy (such as mobility and bodily control) rather than traditional domestic roles (Khalil, 2024).

Overall, the results support hypotheses 1 and 2 outlined in the theoretical framework section (III). Women in consanguineous marriages are more likely than those in non-kin marriages to justify intimate partner violence and are especially more likely to rationalize violence related to patriarchal control over women's autonomy. Specifically, women in consanguineous marriages are significantly more inclined to justify IPV when they demonstrate independence—such as going out or refusing sex—but do not show increased acceptance of violence for domestic failures like burning food, even if it causes material loss. These findings imply that consanguinity correlates with stronger normative acceptance of IPV in areas directly related to controlling women's mobility and bodily autonomy, independent of socioeconomic or demographic factors. This supports the argument that kinship consolidation enhances control over women's autonomy, making IPV more justifiable when gendered expectations are breached. Women who justify IPV appear to use more nuanced reasoning, indicating that attitudes towards IPV are often context-dependent.

**Table 6: Consanguinity and Women's Acceptance of IPV: Propensity Score Matching Estimates**

| VARIABLES | (1)<br>Going out | (2)<br>Neglecting children | (3)<br>Arguing | (4)<br>Refusing intercourse | (5)<br>Burning food |
|---|---|---|---|---|---|
| **ATT:** | | | | | |
| Consanguineous marriage | **0.041****** | **0.024**** | 0.023* | **0.03***** | 0.015 |
| | (0.012) | (0.011) | (0.012) | (0.011) | (0.009) |
| N: Treatment group | 8,803 | 8,780 | 8,771 | 8,659 | 8,747 |
| N: Control group | 5,315 | 5,303 | 5,304 | 5,248 | 5,307 |
| **Diagnostics: Balance quality**[16] | | | | | |
| Mean bias (unmatched) | 14.5 | 14.5 | 14.6 | 14.6 | 14.5 |
| Mean bias (matched) | 2.2 | 1.2 | 1.4 | 1.5 | 2.1 |
| % Bias reduction | 85 | 92 | 90 | 90 | 85.5 |
| Pseudo $R^2$ (matched) | 0.001 | 0.000 | 0.000 | 0.000 | 0.001 |

Note: Standard errors are provided in parentheses: * p<0.1, ** p<0.05, *** p<0.01. The average treatment effects on the treated (ATT) under the PSM model are reported here for the matched group. The dependent variables reflect women's acceptance of intimate partner violence. Each variable is coded as 1 if a woman believes it is justified for a husband to hit his wife for any of the following actions: going out without informing him (going out), neglecting children (neglecting children), arguing with him (arguing), refusing to have sex with him (refusing intercourse), or burning the food (burning food). The variables are coded as 0 if a woman does not agree with the statement. The covariates for matching include women's age, education, husband's education, say in choosing a husband, household wealth, number of household members, residence in rural or urban areas, and region effects. Sample sizes vary slightly across outcomes due to missing responses in individual questions, and are unlikely to affect the main results (less than 2% variation).

Our findings also align with a recent study by Malik et al. (2024), which investigated the gap in IPV justification between women in consanguineous and non-consanguineous marriages in Pakistan using data from the 2017–18 PDHS. They employed logistic regression and Fairlie decomposition analysis to determine how much observable socioeconomic differences explained this attitude gap. Their results indicated that consanguineous marriage significantly raised the likelihood of justifying IPV in three out of five scenarios—specifically, going out without permission (OR = 1.325), arguing with the husband (OR = 1.343), and refusing sexual intercourse (OR = 1.262)—though it did not significantly impact attitudes toward neglecting children or burning food. Younger women, along with those with lower levels of education and wealth, are more likely to justify IPV. Notably, women's empowerment, measured as involvement in household decision-making, emerged as the most effective protective factor, roughly halving the odds of justification across all scenarios. The Fairlie decomposition showed how much of the gap in IPV justification linked to consanguinity could be explained by observable differences in characteristics between the two groups. The proportion of the gap that can be explained varies from 46% (neglecting children) to 90% (arguing with husband). Among the factors explaining this portion, household wealth was the most significant, accounting for 64–77% of the gap depending on the scenario. Women's education was the second most important factor, contributing 15–31%, followed by women's involvement in household decision-making at 8–14%.

[16] Balance diagnostics indicate that the quality matches is high. Mean standardized bias falls from roughly 14.5% in the unmatched sample to 1.2–2.2% after matching (with nearly 85–92% bias reduction), and the pseudo-$R^2$ from the treatment model drops to around 0.000 to 0.001, implying that observed covariates have little remaining power to predict treatment assignment in the matched sample. Thus, conditional on the matched sample, observed characteristics do not systematically differ by marriage type.

Our study extends these insights by asking whether, among women with similar characteristics, being in a consanguineous marriage independently influences acceptance of IPV. Malik et al.'s Fairlie decomposition, meanwhile, investigated how much of the observed gap can be attributed to the fact that women in consanguineous marriages tend to be poorer and less educated. This distinction is crucial because the current study addresses selection bias and seeks to estimate the independent effect of consanguinity, controlling for socioeconomic factors.

**Robustness Checks**

**Inverse Probability Weighting (IPW):** IPW is often used to adjust for confounding in observational studies (Austin and Stuart, 2015; Chesnaye et al., 2022). It reweights the sample so that treatment assignment mimics random conditions conditional on observed covariates, allowing for estimation of causal effects. In this study's context, IPW estimates how women's acceptance of IPV would differ if women in consanguineous marriages had instead been in non-consanguineous marriages, holding observed characteristics constant.

**Kernel matching:** Unlike nearest-neighbor matching, kernel matching does not discard observations and instead constructs counterfactual outcomes for each treated unit using all available control variables. Kernel matching assigns weights to every control observation based on how close its propensity score is to that of a treated observation. Controls with more similar propensity scores receive higher weights; those farther away receive lower weights. Kernel matching improves precision by retaining the full sample while limiting bias through greater weighting of closer propensity score matches (Garrido et al., 2014).

**Radius matching:** Unlike the nearest-neighbor approach, radius matching includes all control observations within a predefined distance (radius) of each treated unit's propensity score. Controls are then weighed according to their proximity, and regression-based bias correction is applied to remove residual imbalances. Huber et al. (2015) show that radius matching with bias adjustment is highly flexible and performs well when the radius is set to a large,data-driven value based on pair matching distances, important confounders are included, and regression adjustment is applied to remove residual bias. We implement this approach using kernel matching with the Epanechnikov kernel, which can be interpreted as radius matching with distance weighting.

To assess the sensitivity of the estimated effects to the choice of matching algorithm, we re-estimate the average treatment effect on the treated (ATT) using alternative propensity score methods, including nearest-neighbor matching with multiple neighbors (Table 7 shows results for 1:3 matching), radius matching, kernel matching, and inverse probability weighting. Across alternative matching methods, the direction and relative magnitude of the estimated effects remain stable. In particular, the most robust effects continue to appear for outcomes related to women's mobility and sexual autonomy, supporting the second hypothesis that acceptance of IPV is more directly linked to control over women's bodily autonomy. In contrast, acceptance of IPV for failing other domestic roles remains relatively smaller and appears statistically insignificant for burning food (despite it being a material and financial loss). The results across alternative matching methods indicate that the main findings are robust.

**Table 7: Consanguinity and Women's Acceptance of IPV: Robustness Checks**

| Scenario | 1:3 NN | Radius | Kernel | IPW |
|---|---|---|---|---|
| Going out without telling husband | **0.035*** ** | **0.036*** ** | **0.035*** ** | **0.036*** ** |
| Neglecting the children | 0.011 | **0.019** ** | **0.017** ** | **0.018** ** |
| Arguing with husband | **0.025** ** | **0.024** ** | **0.023** ** | **0.023*** ** |
| Refusing sexual intercourse | **0.032*** ** | **0.039*** ** | **0.039*** ** | **0.037*** ** |
| Burning the food | 0.002 | 0.011 | 0.011 | 0.011 |

**Notes:** The average treatment effect on the treated (ATT) is reported in percentage points. The treatment here is being in a consanguineous marriage. Estimates based on kernel propensity score matching use the Epanechnikov kernel with the default bandwidth. Propensity scores were estimated via probit regression, including covariates such as age, education, husband's education, say in choosing husband, wealth quintiles, number of members in the household, urban/rural residence, and region effects.
***p<0.01, **p<0.05, *p<0.10.

**Mahalanobis distance matching: Mahalanobis distance matching is** a multivariate matching technique that pairs treated and control units based on their similarity across observed covariates, using a distance metric that accounts for differences in scale and covariance among covariates. Unlike propensity score–based approaches, which rely on a single scalar index, Mahalanobis matching directly minimizes the joint distance between observations in the covariate space, giving greater weight to differences along dimensions with lower variance and appropriately scaling correlated covariates (Mitra et al., 2021). In Table 8, we implement one-to-three nearest-neighbor Mahalanobis matching with replacement to estimate the average treatment effect on the treated, matching women in consanguineous marriages to observationally similar women in non-consanguineous marriages based on the same covariates used in other results above.[17] We also conducted balance tests after matching, which indicated excellent covariate balance. Standardized mean differences for all matched covariates appeared extremely small (between 0.0% and 2.5%)[18], with no statistically significant differences between treated (women in consanguineous marriages) and control groups (women in non-kin marriages). The post-matching pseudo $R^2$ also appeared effectively zero, and joint significance tests failed to reject equality of covariates, suggesting that treatment assignment is no longer predictable from observed characteristics. The results indicated that the matched samples were highly comparable on observed covariates.

Table 8 presents estimates from Mahalanobis matching to assess the robustness of the results under propensity score matching. Overall, the findings remain consistent with Hypothesis 2: consanguineous marriage is associated with higher acceptance of (IPV), particularly in domains that challenge patriarchal control over women's bodily agency. Across four of the five scenarios, women in consanguineous marriages were more likely to justify IPV relative to matched women in non-kin marriages. The largest effects appear in domains related to women's mobility and bodily autonomy, with consanguinity increasing acceptance of IPV by 2.6 percentage points when a wife goes out without permission and by 2.2 percentage points when she refuses sexual intercourse. We observe more modest but still significant effects for neglecting children (1.5 percentage points) and arguing with the husband (1.2 percentage points). In contrast, the effect for burning food, a task linked to routine household performance rather than autonomy or exit threats, is small and statistically insignificant.

[17] We implement Mahalanobis distance matching with three nearest neighbors. The matching distance is computed using Mahalanobis distance on key demographic and household characteristics (age, education, husband's education, say in choosing a husband, household wealth, number of household members, residence in rural or urban areas, and region effects).
[18] A common rule of thumb is that standardized bias below **5%** (or even 10%) indicates good balance.

**Table 8: Consanguinity and Women's Acceptance of IPV: Mahalanobis Matching Estimates**

| | (1) | (2) | (3) | (4) | (5) |
|---|---|---|---|---|---|
| VARIABLES | Going out | Neglecting children | Arguing | Refusing intercourse | Burning food |
| **ATT:** | | | | | |
| Consanguineous marriage | **0.026***** | 0.015* | 0.012* | **0.022***** | 0.005 |
| | (0.009) | (0.009) | (0.010) | (0.009) | (0.008) |
| N: Treatment group | 8,803 | 8,780 | 8,771 | 8,659 | 8,747 |
| N: Control group | 5,315 | 5,303 | 5,304 | 5,248 | 5,307 |

Note: Standard errors are provided in parentheses: * p<0.1, ** p<0.05, *** p<0.0. The table reports the average treatment effect on the treated (ATT) from Mahalanobis distance matching. The dependent variables reflect women's acceptance of intimate partner violence against women in specific scenarios. Each variable is coded as 1 if a woman believes it is justified for a husband to hit his wife for any of the following actions: going out without informing him (going out), neglecting children (neglecting children), arguing with him (arguing), refusing to have sex with him (refusing intercourse), or burning the food (burning food). The variables are coded as 0 if a woman does not agree with the statement. Covariates used for matching include women's age, education, husband's education, say in choosing a husband, household wealth, number of household members, residence in rural or urban areas, and region effects.

Although the Mahalanobis matching estimates in Table 8 are smaller than those from propensity score matching, the overall pattern of effects remains stable, suggesting that women in consanguineous marriages are more likely to justify IPV in specific scenarios than their counterparts in non-kin marriages. The estimates also indicate that acceptance of IPV is context-dependent, with women selectively justifying violence in scenarios that directly challenge male authority, as a means of enforcing patriarchal norms. This pattern is especially prevalent in scenarios related to women's mobility and sexual autonomy, where kinship networks have strong interests in maintaining compliance and restricting women's exit opportunities.

## VI. CONCLUSION

This paper investigates the relationship between consanguineous marriage and women's acceptance of intimate partner violence (IPV) in Pakistan, contributing to a growing literature on how kinship structures shape gender norms and women's agency. Using data from 15,065 ever-married women in the Pakistan Demographic and Health Survey 2017–18, and employing logistic regression, ordered logistic regression, propensity score matching, and Mahalanobis distance matching to ensure robustness, we document several key findings.

First, we show that consanguineous marriages are significantly associated with higher acceptance of IPV across most scenarios. Women's empowerment in the domestic decision-making process — even partial empowerment, defined as participation in at least one household decision-making domain — consistently reduces IPV acceptance across all five scenarios, as does their educational attainment. Crucially, however, the consanguinity coefficient remains stable and significant after controlling for both empowerment measures, confirming that kinship structure operates through a distinct channel that is not captured by individual-level decision-making power alone.

Second, the association between consanguineous marriage and IPV acceptance is not uniform across scenarios but follows a theoretically coherent gradient. Women in consanguineous marriages show the strongest elevated acceptance of IPV in domains that directly challenge patriarchal authority over their bodies and mobility — going out without permission and refusing sexual intercourse — and no significant elevation for burning food, a scenario involving routine domestic performance rather than an assertion of autonomy. This gradient effect appears robust to different estimation methods, specifications, and matching estimators, and holds after controlling for education, household wealth, urban residence, household size, marital choice, past violence experience, and regional fixed effects. The robustness of this pattern supports the interpretation that consanguineous marriage selectively reinforces patriarchal control rather than uniformly endorsing violence against women.

Interpreted through Kandiyoti's (1988) framework of patriarchal bargaining, our findings suggest that women in consanguineous marriages do not indiscriminately endorse IPV. Rather, their attitudes reflect a selective internalization of norms that prioritize male authority precisely in the domains where overlapping natal and marital kin networks have the greatest collective stake in enforcing compliance — women's spatial mobility and bodily autonomy. Consanguineous marriage consolidates extended kinship hierarchies that intensify normative surveillance, constrain women's exit options, and render violence more easily rationalized as legitimate discipline when these gendered expectations are violated. Perceived domestic task failures such as burning food, by contrast, do not threaten the core of the patriarchal bargain in the same way and accordingly attract lower acceptance, even though they represent tangible material losses.

The results also speak to broader debates in the literature on the determinants of IPV acceptance. Women's own education is the most consistent protective factor across all five scenarios. Household wealth shows a strong monotonic gradient, with women in the richest quintile dramatically less likely to justify IPV than those in the poorest. Urban residence, by contrast, is never significant after controlling for education, wealth, and regional fixed effects, suggesting that urbanicity operates primarily through these structural channels rather than independently. The regional fixed effects reveal striking heterogeneity: acceptance rates in FATA exceed those in ICT by over 70 percentage points across several scenarios, underscoring that where a woman lives — and the governance structures, customary legal frameworks, and normative environments that characterize her region — shapes her attitudes toward IPV as powerfully as her individual socioeconomic characteristics. Analyses of IPV acceptance in Pakistan that omit regional fixed effects therefore risk severely misattributing regional variation to individual-level factors.

While our study contributes a new mechanism for understanding women’s attitudes towards the acceptance of IPV, we also acknowledge the limitations of our analysis. Propensity score matching improves comparability on observed characteristics but cannot account for unmeasured confounders — deep-seated social norms, intra-household dynamics, or personality traits that jointly predict both marriage type and IPV attitudes. The estimates should therefore be interpreted as associations conditional on observed characteristics rather than causal effects. The dependent variable captures expressed attitudes rather than experienced violence, and these are related but distinct constructs. Women's responses may also be affected by social desirability bias or interview conditions: Raza and Pals (2025) demonstrate that women overheard by family members during PDHS interviews report substantially higher acceptance than those interviewed privately, and if consanguineous married women are more likely to be under kinship surveillance during interviews, the true attitudinal difference may differ from our estimates. Finally, the cross-sectional design precludes analysis of how attitudes evolve over time or in response to changes in household structure and economic circumstances.

Nevertheless, the policy implications of our findings are consequential at three levels. At the intervention level, the selectivity of the consanguinity effect — concentrated in autonomy-related domains rather than domestic task failures — suggests that IPV prevention programs in high-consanguinity settings should target norms around women's mobility, sexual autonomy, and the right to express disagreement within marriage, rather than relying on generic violence awareness campaigns that treat IPV acceptance as a uniform construct. At the network level, the finding that consanguineous marriages embed women within overlapping kinship hierarchies that enforce patriarchal compliance implies that couple-centered interventions are structurally insufficient. Programs that target only intimate partners leave intact the broader kin network — mothers-in-law, elder women, male relatives, community elders — that typically serves as the primary enforcer of the norms legitimizing violence. Effective prevention in high-consanguinity settings must therefore recruit extended family members, particularly elder women and male kin, as active agents of norm change rather than passive bystanders. At the regional level, the large variation documented here underscores that nationally calibrated IPV programs will systematically underserve women in FATA, KPK, and Balochistan, where acceptance is highest and where the structural conditions sustaining it — limited state reach, customary legal frameworks, lower female education and labor force participation — remain most deeply entrenched. Context-specific program design, grounded in the institutional and normative realities of each region, is not optional but essential.

More broadly, our research establishes that the marriage system — and specifically the kinship structure through which marriages are organized — is an important but largely overlooked institutional determinant of women's attitudes toward IPV. By demonstrating that consanguineous marriage independently and selectively predicts IPV acceptance even among observationally similar women, this research opens a new line of inquiry into the institutional foundations of patriarchal gender norms and their implications for women's agency, voice, and safety within marriage. Future research should examine whether these patterns persist in other high-consanguinity contexts in South Asia and the Middle East, and whether targeted interventions that engage kinship networks can effectively shift the normative frameworks that consanguineous marriage sustains.

**Data Availability:** The dataset, description of the tools, and other supplementary material are available upon request

## References

1. Agha, N. (2016). Kinship in rural Pakistan: Consanguineous marriages and their implications for women. *Women's Studies International Forum*, 54, 1–10.
2. Ahmad, J., Khan, N., & Mozumdar, A. (2021). Spousal violence against women in India: A social–ecological analysis using data from the National Family Health Survey 2015 to 2016. *Journal of Interpersonal Violence*, *36*(21-22), 10147-10181.
3. Akbary, M. F., Ariyo, T., & Jiang, Q. (2022). Sociocultural determinants of attitudes toward domestic violence among women and men in Afghanistan: Evidence from the Afghanistan Demographic and Health Survey 2015. *Journal of Interpersonal Violence*, 37(11–12), NP9321–NP9346.
4. Akram, N. (2021). Household factors forcing women to accept domestic violence in Pakistan. *Violence and gender*, *8*(4), 208-217.
5. Amir-ud-Din, R., Fatima, S., & Aziz, S. (2021). Is attitudinal acceptance of violence a risk factor? An analysis of domestic violence against women in Pakistan. *Journal of Interpersonal Violence*, *36*(7-8), NP4514-NP4541.
6. Anukriti, S., & Dasgupta, S. (2017). Marriage markets in developing countries. In *The Oxford Handbook of Women and the Economy* (pp. 97–120). Oxford University Press.
7. Austin, P. C. (2010). The performance of different propensity score methods for estimating difference in proportions (risk differences or absolute risk reductions) in observational studies. *Statistics in Medicine, 29*, 2137–2148.
8. Austin, P. C. (2014). A comparison of 12 algorithms for matching on the propensity score. *Statistics in Medicine*, *33*(6), 1057-1069.
9. Austin, P. C., & Stuart, E. A. (2015). Moving towards best practice when using inverse probability of treatment weighting (IPTW) using the propensity score to estimate causal treatment effects in observational studies. *Statistics in Medicine*, *34*(28), 3661-3679.
10. Bhatta, D. N., & Haque, A. (2015). Health problems, complex life, and consanguinity among ethnic minority Muslim women in Nepal. *Ethnicity & Health*, *20*(6), 633-649.
11. Biswas, R. K., Rahman, N., Kabir, E., & Raihan, F. (2017). Women's opinion on the justification of physical spousal violence: A quantitative approach to model the most vulnerable households in Bangladesh. *PLOS ONE*, *12*(11), e0187884.
12. Bittles, A. H. (1994). The role and significance of consanguinity as a demographic variable. *Population and Development Review*, 20(3), 561–584.
13. Bittles, A. H. (2012). *Consanguinity in context*. Cambridge University Press.
14. Campbell, O. L. K., Padilla-Iglesias, C., Fiorio, G., & Mace, R. (2024). Genetic markers of cousin marriage and honour cultures. *Evolution and Human Behavior*, 45(6), 106636.
15. Charsley, K. (2007). Risk, trust, gender and transnational cousin marriage among British Pakistanis. *Ethnic and Racial Studies*, 30(6), 1117–1131.
16. Chesnaye, N. C., Stel, V. S., Tripepi, G., Dekker, F. W., Fu, E. L., Zoccali, C., & Jager, K. J. (2022). An introduction to inverse probability of treatment weighting in observational research. *Clinical Kidney Journal*, *15*(1), 14-20.
17. Do, Q.-T., Iyer, S., & Joshi, S. (2013). The economics of consanguineous marriages. *Review of Economics and Statistics*, 95(3), 904–918.
18. Edlund, L. (2018). *Cousin marriage is not choice: Muslim marriage and underdevelopment*. *AEA Papers and Proceedings, 108*, 353–357.
19. Gallup Pakistan. (2025, December 2). *Attitudes toward wife beating in Pakistan: Analysis of PDHS 2017–18*. Gallup Pakistan Big Data Analysis Series. https://gallup.com.pk/wp/wp-content/uploads/2025/12/Gallup-Analysis-of-PDHS-Attitudes-Towards-Wife-Beating-in-Pakistan.pdf
20. Gallup Pakistan. (2024, July 9). *Education attainment by gender: Provincial Census Report Punjab 2017*. https://gallup.com.pk/post/39892
21. Garrido, M. M., Kelley, A. S., Paris, J., Roza, K., Meier, D. E., Morrison, R. S., & Aldridge, M. D. (2014). Methods for constructing and assessing propensity scores. *Health Services Research*, *49*(5), 1701-1720.
22. Gunarathne, L., Apputhurai, P., Nedeljkovic, M., & Bhowmik, J. (2026). Factors influencing attitudes toward intimate partner violence among married women: a multilevel analysis across 20 low-and middle-income countries. *Violence Against Women*, *32*(5), 1039-1055.

23. Haj-Yahia, M. M. (1998). A patriarchal perspective of beliefs about wife beating among Palestinian men from the West Bank and the Gaza Strip. *Journal of Family Issues*, 19(5), 595–621.
24. Hamamy, H. (2012). Consanguineous marriages: Preconception consultation in primary health care settings. *Journal of Community Genetics*, 3, 185–192.
25. Hamamy, H., & Alwan, S. (2016). The sociodemographic and economic correlates of consanguineous marriages in highly consanguineous populations. In D. Kumar & R. Chadwick (Eds.), *Genomics and Society* (pp. 335–361). Academic Press.
26. Huber, M., Lechner, M., & Steinmayr, A. (2015). Radius matching on the propensity score with bias adjustment: tuning parameters and finite sample behaviour. *Empirical Economics*, *49*(1), 1-31.
27. Hussain, S., & Jullandhry, S. (2020, September). Are urban women empowered in Pakistan? A study from a metropolitan city. In *Women's Studies International Forum* (Vol. 82, p. 102390). Pergamon.
28. International Labour Organization. (2019). *Global Wage Report 2018/19*. https://www.ilo.org/resource/news/global-wage-report-201819-what-lies-behind-gender-pay-gap
29. Iqbal, M., & Fatmi, Z. (2021). Prevalence of emotional and physical intimate partner violence among married women in Pakistan. *Journal of Interpersonal Violence*, *36*(9-10), NP4998-NP5013.
30. Jesmin, S. S. (2017). Social determinants of married women's attitudinal acceptance of intimate partner violence. *Journal of Interpersonal Violence*, *32*(21), 3226-3244.
31. Kandiyoti, D. (1988). Bargaining with patriarchy**.** *Gender & Society*, 2(3), 274–290.
32. Kelmemi, W., Chelly, I., Kharrat, M., & Chaabouni-Bouhamed, H. (2015). Consanguinity and homozygosity among Tunisian patients with an autosomal recessive disorder. *Journal of Biosocial Science*, 47(6), 718–726.
33. Khalil, S. (2024). Unpacking the link between cousin marriage and women's paid work. *Social Science Research*, *123*, 103061.
34. Khalil, S. (2025). Gender and neighborhood penalties in Karachi's information technology sector. *Journal of Behavioral and Experimental Economics*, 102469.
35. Khalil, S., & Warner, A. (2025). Invisible labor, visible barriers: The socioeconomic realities of women's work in Pakistan. *arXiv preprint arXiv:2508.16664*.
36. LaBore, K., Ahmed, T., Rizwan-ur-Rashid, & Ahmed, R. (2021). Prevalence and predictors of violence against women in Pakistan. *Journal of Interpersonal Violence*, *36*(13-14), NP7246-NP7263.
37. La Mattina, G., & Shemyakina, O. N. (2024). Growing up amid armed conflict: Women's attitudes toward domestic violence. *Journal of Comparative Economics*, *52*(3), 645-662.
38. Lassi, Z. S., Ali, A., & Meherali, S. (2021). Women's participation in household decision making and justification of wife beating: a secondary data analysis from Pakistan's demographic and health survey. *International Journal of Environmental Research and Public Health*, *18*(19), 10011.
39. Lenze, J., & Klasen, S. (2017). Does women's labor force participation reduce domestic violence? Evidence from Jordan. *Feminist Economics*, 23(1), 1–29.
40. Lunt, M. (2014). Selecting an appropriate caliper can be essential for achieving good balance with propensity score matching. *American Journal of Epidemiology, 179*(2), 226–235.
41. Madhani, F. I., Karmaliani, R., Patel, C., Bann, C. M., McClure, E. M., Pasha, O., & Goldenberg, R. L. (2017). Women's perceptions and experiences of domestic violence: An observational study from Hyderabad, Pakistan. *Journal of Interpersonal Violence*, *32*(1), 76-100.
42. Malik, M. I., Nadeem, M., & Waheed, A. (2024). Consanguineous marriages and the perception of wife-beating justification in Pakistan: An application of Fairlie decomposition analysis. *Journal of Interpersonal Violence*, 39(21–22), 1–27.
43. Makino, M. (2019). Marriage, dowry, and women's status in rural Punjab, Pakistan. *Journal of Population Economics*, 32(3), 769–797.
44. Mitra, A., Bang, J. T., & Abbas, F. (2021). Do remittances reduce women's acceptance of domestic violence? Evidence from Pakistan. World Development, 138, 105149. https://doi.org/10.1016/j.worlddev.2020.105149
45. Mobarak, A. M., Chaudhry, T., Brown, J., Zelenska, T., Khan, M. N., Chaudhry, S., Wajid, R. A., Bittles, A. H., & Li, S. (2019). Estimating the health and socioeconomic effects of cousin marriage in South Asia. *Journal of Biosocial Science*, 51(3), 418–435.
46. Mondal, D., & Paul, P. (2021). Associations of power relations, wife-beating attitudes, and controlling behavior of husband with domestic violence against women in India: Insights from the National Family Health Survey–4. *Violence Against Women*, 27(14), 2530–2551.

47. Nadeem, M., & Malik, M. I. (2021). The role of social norm in acceptability attitude of women toward intimate partner violence in Punjab, Pakistan. *Journal of Interpersonal Violence*, *36*(21-22), NP11717-NP11735.
48. Nasrullah, M., Muazzam, S., Khosa, F., & Khan, M. M. H. (2017). Child marriage and women's attitude towards wife beating in a nationally representative sample of currently married adolescent and young women in Pakistan. *International Health*, *9*(1), 20-28.
49. Pettitt, L. A., Biswas, R. K., & Bhowmik, J. (2024). Women's attitudes toward intimate partner violence in low-and middle-income countries of Southern Asia. *American Journal of Health Promotion*, *38*(1), 12-18.
50. Rahaman, M., Sen, S., Rana, M. J., & Ghosh, S. (2022). Is consanguineous marriage related to spousal violence in India? Evidence from the National Family Health Survey, 2015–16. *Journal of Biosocial Science*, *54*(6), 959-974.
51. Raza, F., & Pals, H. (2025). Attitudes toward wife beating in Pakistan: over-time comparative trends by gender. *Violence Against Women*, *31*(2), 398-420.
52. Rosenbaum, P. R., & Rubin, D. B. (1983). *The central role of the propensity score in observational studies for causal effects. Biometrika*, 70(1), 41–55.
53. Sandridge, A. L., Takeddin, J., Al-Kaabi, E., & Frances, Y. (2010). Consanguinity in Qatar: Knowledge, attitude and practice in a population born between 1946 and 1991. *Journal of Biosocial Science*, 42(1), 59–82.
54. Sardinha, L., & Nájera Catalán, H. E. (2018). Attitudes towards domestic violence in 49 low-and middle-income countries: A gendered analysis of prevalence and country-level correlates. *PLOS ONE*, *13*(10), e0206101.
55. Sarwar, G., & Farid, A. (2024). Unveiling gender disparities in Pakistan: Challenges, progress, and policy implications for achieving SDG 5. *Journal of Development and Social Sciences*, *5*(2), 506-517.
56. Sarwar, A., & Imran, M. K. (2019). Exploring Women's multi-level career prospects in Pakistan: Barriers, interventions, and outcomes. *Frontiers in Psychology*, *10*, 1376.
57. Saud, M., Ashfaq, A., & Mas'udah, S. (2021). Women's attitudes towards wife beating and its connection with intimate partner violence (IPV): An empirical analysis of a national demographic and health survey conducted in Pakistan. *Journal of International Women's Studies*, *22*(5), 149–160.
58. Sayem, A. M., Begum, H. A., & Moneesha, S. S. (2012). Attitudes towards justifying intimate partner violence among married women in Bangladesh. *Journal of Biosocial Science*, *44*(6), 641-660.
59. Shams, M., Kianfard, L., Parhizkar, S., & Mousavizadeh, A. (2020). Women's views about domestic violence: A qualitative study in Iran. *Journal of Interpersonal Violence*, *35*(17-18), 3666-3677.
60. Shenk, M. K., Towner, M. C., Voss, E. A., & Alam, N. (2016). Consanguineous marriage, kinship ecology, and market transition. *Current Anthropology*, 57(S13), S167–S180.
61. Shreemoyee, S., Roychowdhury, P., & Dhamija, G. (2025). Women's attitudes towards physical intimate partner violence in India: Trends, patterns, and determinants. *PLOS ONE*, *20*(3), e0318350.
62. Shrestha, S. K., Thapa, S., Vicendese, D., & Erbas, B. (2022). Women's attitude towards intimate partner violence and utilization of contraceptive methods and maternal health care services: an analysis of nationally representative cross-sectional surveys from four South Asian countries. *BMC Women's Health*, *22*(1), 215.
63. Stuart, E. A. (2010). Matching methods for causal inference: A review and a look forward. *Statistical Science, 25*(1), 1–21.
64. Tadmouri, G. O., Nair, P., Obeid, T., Al Ali, M. T., Al Khaja, N., & Hamamy, H. A. (2009). Consanguinity and reproductive health among Arabs. *Reproductive Health*, 6(1), 1–9.
65. Tayyab, F., Kamal, N., Akbar, T., & Zakar, R. (2017). Men and women's perceptions of justifications of wife beating: Evidence from Pakistan Demographic and Health Survey 2012–13. *Journal of Family Violence*, 32(7), 721–730.
66. Tran, T. D., Nguyen, H., & Fisher, J. (2016). Attitudes towards intimate partner violence against women among women and men in 39 low-and middle-income countries. *PLOS ONE*, *11*(11), e0167438.
67. UN Women. (2016). *Women's economic participation and empowerment in Pakistan*. https://asiapacific.unwomen.org/sites/default/files/Field%20Office%20ESEAsia/Docs/Publications/2016/05/PK-WEE-Status-Report-lowres.pdf
68. Waltermaurer, E. (2012). Public justification of intimate partner violence: A review of the literature. *Trauma, Violence, & Abuse*, *13*(3), 167-175.
69. Weinreb, A. A. (2008). Characteristics of women in consanguineous marriages in Egypt, 1988–2000. *European Journal of Population*, 24, 185–210.

70. World Bank. (2024). *Women, business and the law 2024*. https://documents1.worldbank.org/curated/en/099110624115537381/pdf/P167792-e884b79d-fbaa-4b87-a416-b44e57ac5a89.pdf
71. World Economic Forum. (2025). *Global Gender Gap Report 2025*. https://reports.weforum.org/docs/WEF_GGGR_2025.pdf
72. Yount, K. M., Halim, N., Schuler, S. R., & Head, S. (2013). A survey experiment of women's attitudes about intimate partner violence against women in rural Bangladesh. *Demography*, *50*(1), 333-357.
73. Zaatut, A., & Haj-Yahia, M. M. (2016). Beliefs about wife beating among Palestinian women from Israel: The effect of their endorsement of patriarchal ideology. *Feminism & Psychology*, 26(4), 405–425.
74. Zulfiqar, A., Kuskoff, E., Povey, J., & Baxter, J. (2026). Homemaker or breadwinner: labour force participation of Pakistani women. *Community, Work & Family*, *29*(1), 23-42.

## Appendix

**Table A1: Men and Women's Acceptance of IPV**

| Scenario | Women (N= 3334) | | Men (N=3332) | |
|---|---|---|---|---|
| | **Accept** | Do Not Accept | **Accept** | Do Not Accept |
| Going out without permission | **35.42%** | 62.18% | **27.46%** | 72.21% |
| Arguing with husband | **34.76%** | 62.78% | **20.38%** | 79.38% |
| Refusing sexual intercourse | **29.51%** | 66.68% | **9.66%** | 89.59% |
| Neglecting children | **28.79%** | 68.72% | **18.04%** | 81.51% |
| Burning food | **18.33%** | 79.18% | **3.87%** | 95.8% |

**Notes:** The Pakistan Demographic and Health Survey 2017-18 includes separate samples of 15,068 ever-married women aged 15-49 and 3,691 ever-married men aged 15-49. In this table, sample sizes may vary slightly due to missing responses (<2% variation). "Don't know" responses were excluded (<1% for both samples). Percentages are rounded off.

**Table A2: Women's Acceptance of IPV by Scenario and Region**

| Region | Going out without permission | Neglecting children | Arguing with husband | Refusing sexual intercourse | Burning food |
|---|---|---|---|---|---|
| Punjab | 24.7% *(3,372)* | 23.6% *(3,379)* | 26.3% *(3,374)* | 24.3% *(3,334)* | 15.6% *(3,384)* |
| Sindh | 27.6% *(2,645)* | 26.5% *(2,642)* | 24.7% *(2,647)* | 21.8% *(2,624)* | 17.9% *(2,646)* |
| KPK | 48.8% *(2,293)* | 38.6% *(2,288)* | 55.1% *(2,303)* | 45.7% *(2,267)* | 22.9% *(2,301)* |
| Balochistan | 55.6% *(1,654)* | 39.2% *(1,623)* | 47.0% *(1,610)* | 38.3% *(1,553)* | 21.7% *(1,568)* |
| Gilgit-Baltistan | 46.9% *(972)* | 44.0% *(975)* | 42.6% *(968)* | 39.9% *(970)* | 29.6% *(974)* |
| ICT | 15.2% *(1,102)* | 14.4% *(1,101)* | 15.3% *(1,097)* | 14.6% *(1,086)* | 6.2% *(1,098)* |
| AJK | 22.1% *(1,701)* | 20.4% *(1,703)* | 20.8% *(1,704)* | 18.2% *(1,683)* | 11.2% *(1,704)* |
| FATA | 87.1% *(1,001)* | 58.3% *(996)* | 89.2% *(992)* | 71.4% *(1,005)* | 42.8% *(1,003)* |
| **Total** | **37.1%** *(14,740)* | **30.8%** *(14,707)* | **36.6%** *(14,695)* | **31.6%** *(14,522)* | **19.4%** *(14,678)* |

**Notes:** Cell entries report the share of women who believe a husband is justified in hitting his wife for each scenario listed (this will be referred to as women's acceptance of IPV in this study). Sample sizes are in parentheses. ICT = Islamabad Capital Territory; AJK = Azad Jammu and Kashmir; FATA = Federally Administered Tribal Areas (merged into KPK in 2018).

**Table A3: Consanguineous Marriage, Empowerment, and Women's Acceptance of IPV for Going Out Without Permission (Logistic Regression)**

| Variable | (1) OR [95% CI] | | (2) OR [95% CI] | |
|---|---|---|---|---|
| Consanguineous marriage | **1.523*** ** | *[1.291, 1.797]* | **1.537*** ** | *[1.302, 1.814]* |
| Empowerment (all domains) | 0.938 | *[0.645, 1.366]* | — | — |
| Empowerment (any domain) | — | — | **0.645*** ** | *[0.542, 0.767]* |
| Age | 0.993 | *[0.984, 1.003]* | 0.998 | *[0.988, 1.008]* |
| Respondent's education (years) | **0.940*** ** | *[0.920, 0.960]* | **0.944*** ** | *[0.924, 0.964]* |
| Husband's education (years) | 0.994 | *[0.976, 1.012]* | 0.994 | *[0.976, 1.012]* |
| Wife had a say in choosing husband | 1.052 | *[0.870, 1.273]* | 1.095 | *[0.904, 1.327]* |
| **Wealth index (ref: poorest)** | | | | |
| Poorer | **0.614*** ** | *[0.488, 0.772]* | **0.620*** ** | *[0.493, 0.781]* |
| Middle | **0.402*** ** | *[0.308, 0.524]* | **0.405*** ** | *[0.310, 0.528]* |
| Richer | **0.412*** ** | *[0.304, 0.557]* | **0.418*** ** | *[0.309, 0.567]* |
| Richest | **0.252*** ** | *[0.174, 0.363]* | **0.252*** ** | *[0.174, 0.364]* |
| Number of household members | 1.017 | *[0.995, 1.039]* | 1.011 | *[0.989, 1.033]* |
| Urban residence (ref: rural residence) | 0.924 | *[0.771, 1.109]* | 0.926 | *[0.772, 1.111]* |
| **Past experience with violence** | | | | |
| Experienced any less severe violence | **1.297** ** | *[1.054, 1.596]* | **1.298** ** | *[1.054, 1.598]* |
| Experienced any severe violence | **1.430** ** | *[1.018, 2.008]* | **1.476** ** | *[1.050, 2.073]* |
| Experienced any sexual violence | **1.443* ** | *[0.984, 2.117]* | **1.423* ** | *[0.969, 2.089]* |
| **Region fixed effects (ref: Punjab)** | | | | |
| Sindh | **0.765** ** | *[0.592, 0.988]* | **0.793* ** | *[0.614, 1.025]* |
| KPK | **3.048*** ** | *[2.391, 3.885]* | **2.738*** ** | *[2.139, 3.505]* |
| Balochistan | **2.039*** ** | *[1.539, 2.701]* | **1.910*** ** | *[1.440, 2.534]* |
| Gilgit-Baltistan | **2.335*** ** | *[1.698, 3.211]* | **2.335*** ** | *[1.694, 3.219]* |
| ICT | 0.784 | *[0.527, 1.167]* | 0.793 | *[0.531, 1.183]* |
| AJK | 0.861 | *[0.647, 1.145]* | 0.875 | *[0.658, 1.165]* |
| FATA | **15.398*** ** | *[9.759, 24.295]* | **12.755*** ** | *[8.035, 20.247]* |
| Observations | 3,864 | | 3,864 | |
| Pseudo $R^2$ | 0.2087 | | 0.2135 | |

**Notes:** The dependent variable is a binary indicator equal to 1 if a woman believes a husband is justified in hitting his wife if she leaves the house without informing him, and 0 otherwise. Odds ratios are reported with 95% confidence intervals in brackets. Model (1) includes empowerment across all domains, coded as 1 if the woman participates in all five household decision-making domains, and 0 otherwise. Model (2) includes empowerment in any domain, coded as 1 if she participates in at least one domain. * $p < 0.10$ ** $p < 0.05$ *** $p < 0.01$

**Table A4: Consanguineous Marriage, Empowerment, and Women's Acceptance of IPV for Arguing with Husband (Logistic Regression)**

| **Variable** | **(1)** *OR [95% CI]* | | **(2)** *OR [95% CI]* | |
|---|---|---|---|---|
| Consanguineous marriage | **1.204**** | *[1.021, 1.419]* | **1.209**** | *[1.026, 1.425]* |
| Empowerment (all domains) | 0.762 | *[0.517, 1.122]* | — | — |
| Empowerment (any domain) | — | — | **0.712***** | *[0.599, 0.847]* |
| Age | 1.003 | *[0.993, 1.012]* | 1.006 | *[0.996, 1.016]* |
| Respondent's education (years) | **0.931***** | *[0.911, 0.951]* | **0.933***** | *[0.914, 0.953]* |
| Husband's education (years) | 1.002 | *[0.984, 1.020]* | 1.002 | *[0.984, 1.021]* |
| Wife had a say in choosing husband | 0.983 | *[0.814, 1.189]* | 1.013 | *[0.837, 1.226]* |
| **Wealth index (ref: poorest)** | | | | |
| Poorer | **0.571***** | *[0.453, 0.719]* | **0.576***** | *[0.457, 0.726]* |
| Middle | **0.419***** | *[0.322, 0.547]* | **0.425***** | *[0.325, 0.554]* |
| Richer | **0.389***** | *[0.287, 0.527]* | **0.396***** | *[0.292, 0.537]* |
| Richest | **0.206***** | *[0.142, 0.299]* | **0.208***** | *[0.143, 0.302]* |
| Number of household members | **1.021*** | *[0.999, 1.044]* | 1.017 | *[0.995, 1.040]* |
| Urban residence (ref: rural residence) | 0.998 | *[0.832, 1.197]* | 0.995 | *[0.830, 1.194]* |
| **Past experience with violence** | | | | |
| Experienced any less severe violence | **1.266**** | *[1.028, 1.559]* | **1.265**** | *[1.027, 1.558]* |
| Experienced any severe violence | 1.244 | *[0.887, 1.743]* | 1.276 | *[0.910, 1.788]* |
| Experienced any sexual violence | 1.298 | *[0.880, 1.914]* | 1.287 | *[0.872, 1.899]* |
| **Region fixed effects (ref: Punjab)** | | | | |
| Sindh | **0.648***** | *[0.501, 0.838]* | **0.664***** | *[0.514, 0.859]* |
| KPK | **3.319***** | *[2.607, 4.225]* | **3.071***** | *[2.403, 3.925]* |
| Balochistan | **1.317*** | *[0.995, 1.743]* | 1.258 | *[0.949, 1.667]* |
| Gilgit-Baltistan | **1.450**** | *[1.051, 2.001]* | **1.445**** | *[1.046, 1.997]* |
| ICT | 0.919 | *[0.630, 1.341]* | 0.925 | *[0.633, 1.352]* |
| AJK | **0.779*** | *[0.587, 1.035]* | 0.792 | *[0.596, 1.052]* |
| FATA | **19.025***** | *[11.412, 31.716]* | **16.501***** | *[9.845, 27.658]* |
| Observations | 3,861 | | 3,861 | |
| Pseudo $R^2$ | 0.2083 | | 0.2108 | |

**Notes:** Dependent variable is a binary indicator equal to 1 if a woman believes a husband is justified in hitting his wife if she argues with him, and 0 otherwise. Odds ratios reported with 95% confidence intervals in brackets. Model (1) includes empowerment (all domains), equal to 1 if the woman participates in all five household decision-making domains, and 0 otherwise. Model (2) includes empowerment (any domain), equal to 1 if she participates in at least one domain. * p < 0.10 ** p < 0.05 *** p < 0.01

**Table A5: Consanguineous Marriage, Empowerment (Any Domain), and Women's Acceptance of IPV for Refusing Intercourse, Neglecting Children, and Burning Food (Logistic Regression Results)**

| Variable | (1) Refusing Sexual Intercourse *OR [95% CI]* | | (2) Neglecting Children *OR [95% CI]* | | (3) Burning Food *OR [95% CI]* | |
|---|---|---|---|---|---|---|
| Consanguineous marriage | 1.506*** | *[1.273, 1.782]* | 1.346*** | *[1.144, 1.584]* | 1.056 | *[0.876, 1.273]* |
| Empowerment (any domain) | 0.702*** | *[0.589, 0.838]* | 0.733*** | *[0.616, 0.871]* | 0.622*** | *[0.510, 0.759]* |
| Age | 1.004 | *[0.994, 1.014]* | 1.001 | *[0.991, 1.010]* | 1.003 | *[0.992, 1.014]* |
| Respondent's education (years) | 0.942*** | *[0.921, 0.963]* | 0.932*** | *[0.912, 0.953]* | 0.923*** | *[0.897, 0.949]* |
| Husband's education (years) | 1.009 | *[0.991, 1.028]* | 0.988 | *[0.970, 1.006]* | 0.985 | *[0.965, 1.006]* |
| Wife had a say in choosing husband | **0.798**** | *[0.659, 0.965]* | 1.482*** | *[1.222, 1.797]* | **1.212*** | *[0.977, 1.502]* |
| **Wealth index (ref: poorest)** | | | | | | |
| Poorer | 0.632*** | *[0.504, 0.792]* | **0.711***** | *[0.572, 0.884]* | 0.633*** | *[0.502, 0.798]* |
| Middle | 0.439*** | *[0.337, 0.573]* | 0.516*** | *[0.399, 0.668]* | 0.430*** | *[0.321, 0.575]* |
| Richer | 0.395*** | *[0.290, 0.538]* | 0.487*** | *[0.360, 0.657]* | 0.405*** | *[0.285, 0.575]* |
| Richest | 0.200*** | *[0.136, 0.295]* | 0.276*** | *[0.190, 0.401]* | 0.162*** | *[0.098, 0.269]* |
| Number of household members | 1.015 | *[0.993, 1.038]* | 1.012 | *[0.990, 1.034]* | 1.009 | *[0.984, 1.034]* |
| Urban residence (ref: rural residence) | 0.924 | *[0.768, 1.111]* | 1.017 | *[0.848, 1.220]* | 0.880 | *[0.711, 1.089]* |
| **Past experience with violence** | | | | | | |
| Experienced any less severe violence | **1.223*** | *[0.995, 1.503]* | 0.874 | *[0.711, 1.075]* | **0.732**** | *[0.576, 0.931]* |
| Experienced any severe violence | **1.435**** | *[1.021, 2.016]* | 0.983 | *[0.697, 1.386]* | 0.977 | *[0.656, 1.456]* |
| Experienced any sexual violence | **1.382*** | *[0.946, 2.021]* | **1.937***** | *[1.334, 2.811]* | **1.647**** | *[1.090, 2.489]* |
| **Region fixed effects (ref: Punjab)** | | | | | | |
| Sindh | 0.483*** | *[0.367, 0.635]* | 0.881 | *[0.683, 1.135]* | **0.759*** | *[0.564, 1.021]* |
| KPK | 2.058*** | *[1.604, 2.641]* | 2.007*** | *[1.566, 2.574]* | 1.109 | *[0.826, 1.489]* |
| Balochistan | 1.216 | *[0.912, 1.622]* | 1.000 | *[0.747, 1.339]* | **0.687**** | *[0.487, 0.971]* |
| Gilgit-Baltistan | **1.628***** | *[1.179, 2.249]* | 1.869*** | *[1.355, 2.578]* | **1.369*** | *[0.954, 1.964]* |
| ICT | 0.756 | *[0.500, 1.142]* | **0.642**** | *[0.419, 0.984]* | 0.652 | *[0.376, 1.131]* |
| AJK | **0.670***** | *[0.498, 0.901]* | 0.802 | *[0.599, 1.072]* | 0.816 | *[0.580, 1.148]* |
| FATA | 3.600*** | *[2.520, 5.143]* | 2.738*** | *[1.954, 3.837]* | 1.938*** | *[1.358, 2.765]* |
| Observations | 3,817 | | 3,856 | | 3,866 | |
| Pseudo $R^2$ | 0.1793 | | 0.1339 | | 0.1431 | |

**Notes:** Dependent variables are binary indicators of women's acceptance of intimate partner violence in each scenario. Odds ratios reported with 95% confidence intervals in brackets. All models include empowerment (any domain), equal to 1 if the woman participates in at least one of five household decision-making domains (own healthcare, large purchases, visits to relatives, husband's earnings, own earnings), and 0 otherwise. * $p < 0.10$ ** $p < 0.05$ *** $p < 0.01$